\documentclass{aa}
\usepackage{natbib}
\usepackage{color}
\usepackage{graphicx}
\bibpunct{(}{)}{;}{a}{}{,}
\usepackage{txfonts}
\newcommand{\hd}{HD~112244 }
\newcommand{\hde}{HD~112244}

\newcommand{\spefo}{{\tt SPEFO} }
\newcommand{\spefoe}{{\tt SPEFO}}
\newcommand{\phoebe}{{\tt PHOEBE} }
\newcommand{\phoebee}{{\tt PHOEBE}}
\newcommand{\fotel}{{\tt FOTEL} }

\newcommand{\korel}{{\tt KOREL} }
\newcommand{\korele}{{\tt KOREL}}
\newcommand{\kms}{km~s$^{-1}$ }
\newcommand{\ks}{km~s$^{-1}$}
\newcommand{\vsin}{$v\sin i$ }
\newcommand{\vsi}{$v\sin i$}
\newcommand{\vra}{$v_1\sin i$ }
\newcommand{\vrb}{$v_2\sin i$ }
\newcommand{\vrc}{$v_3\sin i$ }

\newcommand{\tef}{$T_{\rm eff}$ }
\newcommand{\teff}{$T_{\rm eff}$}
\newcommand{\lgg}{{\rm log}~$g$ }
\newcommand{\ms}{M$_{\odot}$}
\newcommand{\rs}{R$_{\odot}$}

\newcommand{\hp}{\hbox{$H_{\rm p}$}}
\newcommand{\oc}{\hbox{$O\!-\!C$}}
\newcommand{\m}{$.\!\!^{\rm m}$}
\newcommand{\nn}{\accent'27A}

\newcommand{\ha}{H$\alpha$ }
\newcommand{\hae}{H$\alpha$}

\begin{document}

\title{Physical properties of seven binary and higher-order multiple OB~systems
\thanks{Based on spectra from observations made with ESO
telescopes at La Silla and Paranal Observatories under programs
065.1-0526(A), 073.C-0337(A), 074.D-0300(A), 075.D-0103(A), 077.D-0705(A),
077.B-0348(A), 079.D-0564(A), 081.D-2008(A), 083.D-0589(A), 085.D-0262(A),
086.D-0997(A), 087.C-0012(A), 087.D-0946(A), 089.C-0415(A), 089.D-0097(B),
089.D-0975(A), 090.C-0280(A), 091.D-0145(A), 095.A-9032(A), and 178.D-0361(E),
program TYCHO P2, on the BESO spectra and spectra from CTIO, and on
the Hipparcos \hp\ and ASAS3 $V$ photometry.}
}

\titlerunning{Seven early-type binary and multiple systems}

\author{Pavel Mayer\inst{1}\and
    Petr Harmanec\inst{1}\and
    Rolf Chini\inst{2,3}\and
    Anita~Nasseri\inst{2}\and
    Jana A. Nemravov\'{a}\inst{1}\and
    Horst~Drechsel\inst{4}\and
    Rodrigo~Catalan-Hurtado\inst{5}\and
    Brad~N.~Barlow\inst{5}\and
    Yves~Fr\'emat\inst{6}\and
    Lenka~Kotkov\'a\inst{7}
}
\institute{
Astronomical Institute of the Charles University,
Faculty of Mathematics and Physics,\\
V~Hole\v{s}ovi\v{c}k\'ach~2, CZ-180 00 Praha~8,
Czech Republic
\and
Astronomisches Institut,
Ruhr--Universit\"at Bochum,
Universit\"atsstr. 150,
44801 Bochum, Germany
\and
Instituto de Astronom{\'i}a,
Universidad Cat\'{o}lica del Norte,
Avenida Angamos 0610,
Casilla 1280, Antofagasta, Chile
\and
Dr.~Karl~Remeis-Observatory \& ECAP, Astronomical Institute,
Friedrich-Alexander-University Erlangen-Nuremberg,
Sternwartstr.~7, 96049 Bamberg, Germany
\and
Department of Physics, High Point University,
One University Way, High Point, NC 27268, USA
\and
Royal Observatory of Belgium, Avenue circulaire 3, B-1180 Brussels, Belgium
\and
Astronomical Institute, Academy of Sciences of the Czech Republic,
CZ-251~65 Ond\v{r}ejov, Czech Republic
}

\authorrunning{P. Mayer et al.}

\date{Received \today; accepted}

% \abstract{}{}{}{}{}
% 5 {} token are mandatory

\abstract{Analyses of multi-epoch, high-resolution ($R \sim 50000$) optical
spectra of seven early-type systems provided various important new insights
with respect to their multiplicity.
First determinations of orbital periods were made for HD~92206~C (2\fd022),
HD~112244 (27\fd665), HD~164438 (10\fd25), HD~123056~A ($\sim 1314$\,d) and
HD~123056~B ($< 2$\,d); the orbital period of HD~318015 could be improved(23\fd445975).
Concerning multiplicity, a~third component was discovered for HD~92206~C by
means of  \ion{He}{i} line profiles.
For HD~93146~A, which was hitherto assumed to be SB1, lines of a secondary
component could be discerned.
HD~123056 turns out to be a multiple system consisting of a high-mass
component~A ($\approx$ O8.5) displaying a~broad \ion{He}{ii}~5411~\AA\
feature with variable radial velocity, and of an~inner pair~B ($\approx$ B0)
with double \ion{He}{i} lines.
The binary HD~164816 was revisited and some of its system parameters were improved.
 In particular, we determined its systemic velocity to
be $- 7$\,km\,s$^{-1}$, which coincides with the radial velocity of
the cluster NGC\,6530. This fact, together with its distance, suggests
the cluster membership of HD~164816.
The OB system HD~318015 (V1082~Sco) belongs to the rare class of eclipsing
binaries with a supergiant primary (B0.5/0.7). Our combined orbital and
light-curve analysis suggests that the secondary resembles an O9.5\,III star.
Our results for a limited sample corroborate the findings that many O stars
are actually massive multiple systems.
%These findings provide further evidence that most high-mass stars are compact
%hierarchical systems. The tendency that the secondary and tertiary components
%are often high-mass stars too will change the initial mass function of open
%clusters and strongly suggests that the multiplicity originates from the star
%formation process rather than from tidal capture.
}
\maketitle
%
%________________________________________________________________

\section{Introduction}

Duplicity or even multiplicity of O stars is a topic that has attracted
attention from various aspects, mainly related to the origin and evolution
of massive stars. While  \citet{garmany80}, for example,
list only 40 O galactic-type binaries with known orbits in the survey
of the binary frequency of O stars, this fraction has been growing steadily.
\citet{mason98} obtained speckle interferometry of 227 O stars brighter
than $V=8$\m0 and compiled information about their spectroscopic duplicity
or multiplicity from the literature. They concluded that the duplicity is higher
 for stars in clusters
and associations (72~\%) than for the field stars
(19~\%); the lowest being for runaway stars (8~\%). They also noted that
many O stars are actually triple systems.  In continuation of this study, \citep{mason09}, they further
 strengthened the multiplicity among clusters
and associations.  \citet{sana2008} re-investigated the duplicity of
O~stars in the NGC~6231 cluster and concluded that more than 63~\% of O stars
must be binaries. They pointed out that this observational
fact represents an important constraint for the theories of early
evolution of O stars.
\cite{chini} studied a sample of 248 O and 581 mostly southern stars and
concluded that more than 82~\% of stars more massive than 16~\ms\
are close binary systems and that this fraction decreases to 20~\% for
3~\ms\ stars. They also found differences in the binary fractions of O stars
in different environments: $72\pm13$~\% in clusters, $73\pm8$~\% in
associations, $69\pm11$~\% among runaway stars and $43\pm13$~\% for the
field stars.
 The fact that duplicity must be a key factor affecting the evolution of
high-mass binaries was discussed in detail by \citet{sana2012}.
In several recent studies the binary nature of many O-type stars was found or suspected.
\cite{sota2014} presented a survey of 448 Galactic O stars from both hemispheres,
based on high signal-to-noise (S/N) ratio $R \sim 2500$ spectra, which is complete to $B = 8$\m0;
further 142 stars were listed by \citet{sota16}. Combining results from a multi-epoch, high-resolution
spectroscopic survey, OWN \citep{barba2010}, with additional information on spectroscopic and visual binaries
from the literature, the authors conclude that only a very small fraction of stars
 with masses above $15 - 20$\,\ms\
are born as single systems.

In the ongoing multi-epoch survey started in 2009 \citep{chini}, which
meanwhile comprises more than 2000 high-resolution spectra, a~large number
of hitherto unknown spectroscopic binaries was found. An example of
the~study where the spectra of this survey were analysed in detail
is HD~152246, discovered as a~binary by \citet{thack}. The analysis of its
double-lined spectra yielded a~significant difference of the orbital periods
for both components. It suggests that HD~152246 is at least a~hierarchical
triple system consisting of an inner pair Ba/Bb with approximately 17 and 6\,\ms,
respectively, and an outer component A with a mass of approximately 20\,\ms\ \citep{Nasseri2014}.

 There are, of course, other studies of individual binaries where
the orbit of a third body is measured interferometrically \citep[]{mayer2014}
and of members of
associations \citep{kiminki2007, kiminki2012, kobul2014, sana2011, sana2013}.

For more details on the subject, we refer readers to a study by
\citet{aldo2015}, which represents a~very good overview of the current
status of duplicity of O stars and also gives a number of relevant references.
The authors identified new binaries with the Hubble Space Telescope fine
guidance sensor and carried out a very extended literature search that allowed them to provide an
estimate of the period distribution
among O~type binaries.
They concluded that it is approximately flat and discussed several selection
effects and the need for additional data. However, their complete
list contains only 83 spectroscopic and eclipsing binaries with known
orbital period.

The present paper brings into focus seven early-type systems, for which
we analysed between 18 and 98 multi-epoch spectra, obtained mainly
with the BESO and FEROS spectrographs.
In addition to new orbital periods, our study reveals higher multiplicities
in some of the systems. Especially important in this context are systems
with a supergiant component (HD~318015, HD~112244), which are still
rare. Reliable determination of their basic physical properties
provides a sensitive test for the models of stellar structure and evolution,
especially those that also include mass loss and/or rotation.
 We admit that our selection of objects was mainly guided by the
above-mentioned practical considerations. It is clear, however,
\citep[cf.][]{aldo2015} that our contribution is not as modest as it might
appear at first sight. So far, the knowledge of duplicity and/or multiplicity
of OB stars has been mainly documented on statistical grounds while
the orbital solutions are still missing in the majority of cases.
\citet{zinnecker} and \citet{sana-evans}, for instance, pointed out that their knowledge is
 important for the understanding of the binary
origin and evolution.

\section{Data}

The data comprise two categories; those that were secured by
our own observations and those downloaded from {the ESO archive} to complement
the phase coverage of the individual systems.

\subsection{Spectrographs}

The majority of spectra were secured with the echelle spectrograph BESO \citep[$R \sim 48000$,][]{BESO}
 of the Universit\"ats-Sternwarte Bochum, which is
fibre-fed to the 1.5~m Hexapod telescope at Cerro Armazones, Chile; the corresponding spectra are
denoted as ``B''. Further spectra were downloaded from the ESO archive. Those ESO spectra that
were secured with the echelle spectrograph FEROS \citep[$R \sim 48000$][]{feros} attached to the
 MPI~2.2~m telescope are labelled ``F'', the spectra from HARPS \citep{harps} mounted at the
3.6~m ESO telescope are labelled ``H''. Some of the FEROS spectra were obtained in the framework
 of the observational program ``Tycho Brahe" allocated to Czech astronomers; these spectra are
denoted as ``T''. In the case of V1082~Sco = HD~318015, we also used the spectra obtained at CTIO.
 The CTIO spectra were secured with the echelle spectrograph CHIRON \citep[$R \sim 28.000,$][]{chiron}
attached to the 1.5~m telescope and are labelled ``C''.
The initial reductions (bias subtraction, flatfielding and wavelength
calibration) were carried out using the standard pipelines for individual instruments.
 Rectification of the spectra and smaller cosmics and flaw-removal were carried out
 interactively using the program \spefo
\citep{sef0, spefo}, the latest version developed by the late Mr.~J.~Krpata.
This program allows the user to define the continuum points on the computer
screen for one well-exposed spectrum, through which the continuum is smoothly
defined by Hermite polynomials. The user can see the fit and is able to
modify it when necessary. When proceeding to the text spectrum, all previously
chosen rectification points are copied to it and the user can modify their
positions if necessary, depending on the noise and radial-velocity (RV) shifts.
We also carefully compared the rectification of each spectrum with the first
(template) spectrum to be sure about uniformity of the rectification.
Similarly, the removal of cosmics can be carried out interactively on
the computer screen, the deleted pixels being replaced by a spline
interpolation. Basic information about all spectra is in Table~\ref{basic}.

\begin{table*}
\caption{Basic information about the spectra used.}
\label{basic}
\begin{tabular}{rccrcrrrc}
\hline\hline\noalign{\smallskip}
Spectrograph&Abbreviation&Spectral Range (\AA)&Resolution& S/N range\\
\hline\noalign{\smallskip}
FEROS       & F plus T   & 3600 -- 9100       &    48000 & 63 -- 311\\
HARPS       &    H       & 3800 -- 6912       &   115000 & 84 -- 106\\
BESO        &    B       & 3800 -- 8000       &    48000 & 26 -- 196\\
CTIO        &    C       & 4500 -- 8900       &    28000 & 83 -- 177\\
\hline\noalign{\smallskip}
\end{tabular}
\end{table*}

\subsection{Radial-velocity and orbit determination}

Depending on the character of spectra of individual objects,
we alternatively applied three different techniques of RV
measurements:

\begin{enumerate}
\item Fit the line profiles by one, two or three Gaussian profiles.
\item Interactive comparison of the direct and flipped line profiles displayed
on the computer screen using the program \spefoe.
\item Two-dimensional (2D) cross-correlation using a Fortran implementation of the {\tt TODCOR}
\citep{todcor1} called {\tt asTODCOR}, developed by one of the present authors (YF) and
first applied to the study of AU~Mon \citep{Desmet}.
\end{enumerate}

%% PM
Whenever using the first two techniques, we preferred spectral lines
with no obvious blends and those strong enough to give reliable results even
for spectra with lower signal to noise ratio (S/N).
Throughout this paper, we record the times of observations using
\emph{reduced} heliocentric Julian dates

\smallskip
\centerline{RJD = HJD - 2400000.0\,.}

\smallskip
The orbital solutions were obtained with the program \fotel \citep{fotel2}
and the combined orbital and light-curve solution for the eclipsing binary
V1082~Sco was obtained with
the latest version of program \phoebe\,1 \citep{prsa2005,prsa2006}. In a few cases,
we also used the program \korel for spectra disentangling \citep{korel3}.
%% PM
{Having the Gaussian fits we could convert their widths to projected
rotational velocities \vsi. There are several studies that provide such
conversion formulae for specific spectral lines. A suitable one for us
is by \citet{munari} for the \ion{He}{i}~5876~\AA\ line. We prefer this
particular line despite the fact that it is contaminated by
several telluric lines since the blue parts of spectra are often noisier.
This is particularly acute for the BESO spectra since the
sensitivity (due to the optical fibre
 used at the time of our observations)
dropped rapidly towards blue wavelengths.

%%PM In the following sections, we discuss the individual sources
%%in more detail.

\section{HD~92206~C}
The star CPD$ -57\degr3580$ (CD$ -57\degr 3378$, LS~1695) is often referred to as
HD~92206~C, although it is not named so in WDS or SIMBAD. It is the second
brightest member of the open cluster NGC~3324; the brightest member being
HD~92206~AB. The open cluster is embedded in the H~II region G~31. These two
cluster members are separated by 36\arcsec\ and surrounded by groups of
fainter stars. HD~92206~C has $V = 9$\m05 and a spectral type
O8\,V + O9.7\,V \citep[according to][]{sota2014}. The SB2 nature of this star
was first reported by \citet{chini}; no RVs had been published before.

\begin{table*}
\caption{RVs of HD~92206~C measured via Gaussian fits of the line profiles.
}
\label{RV92206}
\begin{tabular}{ccrrrrrrc}
\hline\hline\noalign{\smallskip}
RJD       & Phase   &\multicolumn{3}{c}{5411 primary}&
\multicolumn{3}{c}{5876 secondary}&Spg.\\
          &         & RV      &error&\oc     & RV    &error&\oc     \\
\hline\noalign{\smallskip}
54209.5862 &0.2537   & $-213.5$&  8  &   5.7  &   278 &    7 &$-12.3$& F   \\
54209.5979 &0.2595   & $-212.5$&  8  &   6.4  &   277 &    7 &$-12.8$& F   \\
54246.5539 &0.5313   &    19.5 & 11  &$-11.4$ &       &      &       & F   \\
54246.5627 &0.5357   &    32.0 &  8  &$ -4.5$ &       &      &       & F   \\
54247.5640 &0.0321   & $ -52.5$& 17  &$ -0.6$ &       &      &       & F   \\
54600.5388 &0.5550   &    56.5 & 17  &$ -4.4$ & $-144$&   14 &$-47.4$& F   \\
54601.5493 &0.0546   & $ -95.0$& 14  &$-14.7$ &   138 &   20 &  39.6 & F   \\
55016.4880 &0.2180   & $-221.0$& 11  &$ -6.4$ &   304 &   17 &  19.5 & B   \\
55251.7260 &0.5283   &    30.5 &  8  &   3.4  &       &      &       & B   \\
55252.7361 &0.0276   &  $-32.0$& 14  &  14.1  &       &      &       & B   \\
55255.7089 &0.4976   &  $-13.0$&  8  &   0.1  &       &      &       & B   \\
55256.7212 &0.9981   &     3.0 & 14  &  10.5  &       &      &       & B   \\
55289.6709 &0.2896   & $-208.5$& 11  &   4.3  &   270 &    7 &$-11.4$& B   \\
55294.7294 &0.7908   &   206.5 & 11  &  14.0  & $-300$&   11 &$-21.6$& B   \\
56067.5121 &0.8802   &   137.5 &  8  &   4.4  & $-171$&    5 &  25.3 & F   \\
56687.8253 &0.6038   &   120.0 &  8  &   2.9  & $-162$&   14 &  12.2 & B   \\
56730.6903 &0.7978   &   190.0 &  8  &   0.0  & $-266$&    7 &   8.9 & B   \\
56740.6454 &0.7201   &   206.0 &  8  &  10.4  &       &      &       & B   \\
\hline\noalign{\smallskip}
\end{tabular}

All RVs are in \ks. The last column ``Spg.'' denotes the spectrograph.
\end{table*}

There are ten BESO spectra of HD~92206~C; together with eight FEROS spectra
 from the ESO archive, they are listed in Table~\ref{RV92206}. RVs of both
components were measured by Gaussian fitting in the \ion{He}{i}~5876~\AA\
line, while only the primary is also visible in the \ion{He}{ii}~5411~\AA\
line. Therefore, the RVs measured for the latter line appear more representative
for the primary RV curve. Narrow nebular lines are present in the Balmer as well
as in the \ion{He}{i} lines; due to the larger diameter of the BESO fiber,
these emission lines are more pronounced there. Despite the low number of spectra,
it was possible to find a period of 2\fd0225. In the \ion{He}{i} profiles,
 a third component is also clearly present, see Fig.~\ref{TRI}; this component
was not detected by \citet{mason09}.
As the diameter of the fiber in FEROS is 2\arcsec, the component separation should
be smaller than 1\arcsec. There are indications that the RVs of this component are
not constant. However, the superimposed nebular emission and a~low S/N of some
spectra prevent us from finding the period of the third body.

Orbital elements of the close orbit are listed in Table~\ref{orb92206} while the
RV curve is shown in Fig.~\ref{C_92}. In the epoch before RJD $\approx 55100,$ the
negative O-Cs prevail; after RJD, the positive ones are dominant. This might be
a manifestation of the mutual orbit of the binary and the third body.

The equivalent widths (EWs) of the components in the \ion{He}{i}~4922~\AA\ line are
0.27 (primary), 0.21 (secondary) and 0.05~\nn\, (tertiary); the third component
seems to be 2 -- 3 mag fainter than the close binary suggesting a spectral type of B2.
% PM
Assuming reasonable masses and radii, one could speculate that the objectmight be an
ellipsoidal or even an eclipsing binary. The latter possibility is less likely,
however, since no significant Rossiter-McLaughlin effect was detected.
Also, the observed \vsin seems to require a rather high orbital inclination
in case of spin-orbital synchronisation, which is common among very close systems.

Recently, multicolour photometry of the object was obtained by
\citet{hack2015} and Dr.~Hackstein kindly provided us with individual
observations. Regrettably, only infrared magnitudes were not saturated.
They have sufficient phase coverage, however, to exclude the presence
of the binary eclipses. Further, they might indicate some low-amplitude physical
variations, possibly associated with the rotational period of the primary,
but their observational frequency was one observation per night. Therefore,
new dedicated whole-night series of observations would be needed to
(dis)prove the suspected physical light changes.

\begin{figure}
\resizebox{\hsize}{!}{\includegraphics{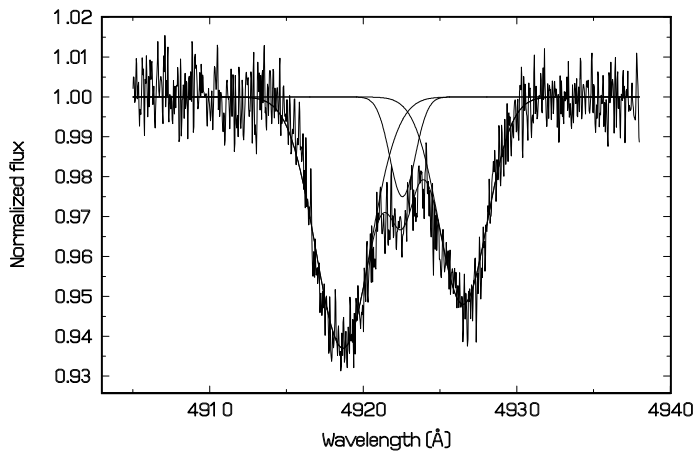}}
\caption{The Gaussian profiles of the primary, secondary and third components in
the FEROS spectrum of HD~92206~C taken on RJD 54209.}
\label{TRI}
\end{figure}

\begin{figure}
\resizebox{\hsize}{!}{\includegraphics{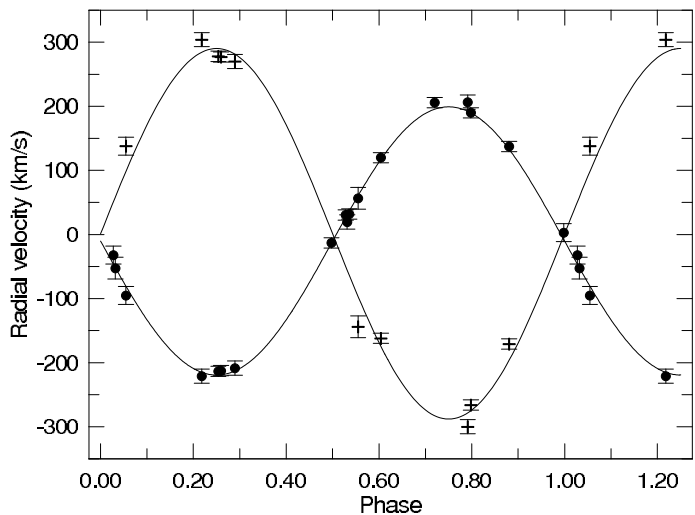}}
\caption{The orbital RV curve of HD~92206~C for the 2\fd0225 orbit.}
\label{C_92}
\end{figure}

\begin{table}
\begin{center}
\caption{Orbital elements for HD~92206~C. All elements have their usual
meaning; rms is the rms of one observation of unit weight derived
from the difference between the calculated and observed RVs.}
\label{orb92206}
%\begin{minipage}
\begin{tabular}{lr}
\hline\hline\noalign{\smallskip}
Element                             &  Value             \\
\hline\noalign{\smallskip}
$P$ [days]                          &     2.022504\,(12) \\
$T_{\rm super. conj.}$ [RJD]        & 55901.9142\,(67)   \\
$e$                                 &     0 (fixed)      \\
$K_1$ [\ks]                         &   209.3\,(4.6)     \\
$K_2$ [\ks]                         &   289.1\,(7.8)     \\
$\gamma_1$ [\ks]                    &    -10.2\,(2.2)    \\
$\gamma_2$ [\ks]                    &     1.2\,(7.8)     \\
$m_1 \sin^3 i$ [\ms]                &    15.0\,(6)       \\
$m_2 \sin^3 i$ [\ms]                &    10.9\,(3)       \\
$a\,\sin i$ [\rs]                   &    21.8\,(4)       \\
\vra [\ks]                          &  146\,(30)         \\
\vrb [\ks]                          &  120\,(5)          \\
\vrc [\ks]                          &   67\,(8)          \\
rms$_1$ [\ks]                          &     8.0            \\
rms$_2$ [\ks]                          &      24            \\
\hline\noalign{\smallskip}
\end{tabular}
\end{center}
\end{table}

\section{HD~93146~A}

HD~93146~A (CPD$ -59\degr2555$, CD$ -59\degr 3278$, LS~1826; $V = 8.44$) is a member of the
open cluster Collinder~228 (No. 65) in the Carina complex at a distance of approximately
 2.3\,kpc; the latter value depends on the adopted reddening law. \citet{Levato1981}
determined a spectral type of O6\,V and noted the variability of RVs; \citet{wal85}
 suggested classification O6.5\,V from IUE spectra. Recently, \citep[][$R \sim 2500$]{sota2014}
 classified the star as O7\,V((f))z. In our spectra the \ion{He}{ii}~4541~\AA\ line is clearly
stronger than the \ion{He}{i}~4471~\AA\ line, so the spectral type O6.5\,V appears more
appropriate. With this classification, the minimum mass of the primary
(see Table~\ref{orb93146}) is also in better agreement with the expectation \citep{martins}.
However, we note that the \ion{He}{i} lines are partly filled by the secondary component,
 which should be taken into account when classifying the spectral type. We also note that the
 Washington Double Star Catalogue \citep[WDS][]{Mason01} lists a nearby (6\farcs5) star
 that is 1.5\,mag fainter than HD~93146~A. \citet{sota2014} use the designation HD~93146\,B
and determine a spectral type of O9.7\,IV. It is not clear at this point if the two stars
form a physically bound system.

\citet{Levato1990} published ten RVs, some of which vary by 25\,\kms from one day to another.
In contrast, \citet{sota2014} reported that OWN data suggest a period of 1130~days.
 We analysed seven BESO and twelve FEROS spectra and derived RVs using both
the Gaussian fitting and {\tt asTODCOR}; examples of the line profiles are
in Fig.~\ref{PriSec}. Following the experience of \citet{zarf30}, we first
disentangled the spectra using \korel and made the preliminary orbital solution
based on the RVs from the Gaussian fitting of line profiles. Then we used the
disentangled spectra as templates for {\tt asTODCOR}. A new version~(R3)
of {\tt asTODCOR} was used, which also provides error estimates of
individual RVs. According to the Appendix in \citet{tridcor1}, the error estimate
of the component RVs derived from multi-dimension cross-correlation
techniques can be obtained from a one dimensional (1D) error analysis carried out on each
dimension. In this version of the program, we therefore applied
a~relation deduced by \citet{zucker2003} from the maximum-likelihood theory that
provides reliable estimates of the uncertainty as long as the errors
are dominated by random processes.

With both methods we could derive RVs also for the secondary component from
the \ion{He}{i}~5876~\AA\ line by fitting the Gaussians, and in the spectral
region from 4460 to 4723~\AA\ using {\tt asTODCOR}.
The spectra and RVs measured by both techniques are listed in
Table~\ref{RV93146}.
 For the orbital solution, we used the {\tt asTODCOR} RVs individually
weighted by weights inversely proportional to the square of their
estimated errors. The orbital RV curve is shown in Fig.~\ref{RV_93146}.
The orbital solution is in Table~\ref{orb93146} and corroborates
a~long period of 1131.6~days.

The EWs of both components for the \ion{He}{i}~5876~\AA\ line are 0.55 and 0.09~\AA,
respectively, which implies a~spectral class of approximately B0.5 for the secondary.
If the visible O9.7\,IV neighbour (HD~93146\,B) were indeed physically bound,
HD~93146~A would be another hierarchical triple system of high-mass stars.

\begin{table*}
\caption{RVs of HD~93146~A measured via Gaussian profiles and using
{\tt asTODCOR} over the spectral region from 4460 to 4723~\AA.}
\label{RV93146}
\begin{tabular}{ccrrrrc}
\hline\hline\noalign{\smallskip}
         \multicolumn{1}{c}{RJD}
       & \multicolumn{1}{c}{Phase}
       & \multicolumn{1}{c}{5876}
       & \multicolumn{1}{c}{5411}
       & \multicolumn{2}{c}{asTODCOR}
       & \multicolumn{1}{c}{Sp.}\\
         \multicolumn{1}{c}{}
       & \multicolumn{1}{c}{}
       & \multicolumn{1}{c}{prim.}
       & \multicolumn{1}{c}{prim.}
       & \multicolumn{1}{c}{prim.}
       & \multicolumn{1}{c}{sec.}
       & \multicolumn{1}{c}{}\\
\hline\noalign{\smallskip}
53363.8162&0.0887&$-18.7$&$-22.9$&$-28.62\pm0.18$&$ 37.12\pm0.22$&F\\
54209.6500&0.8389&  24.6 &  24.8 &$ 24.14\pm0.13$&$-44.28\pm0.16$&F\\
54248.5217&0.8734&  26.5 &  27.0 &$ 25.81\pm0.12$&$-46.31\pm0.15$&F\\
54600.5969&0.1856&$-20.4$&$-20.3$&$-22.98\pm0.10$&$ 41.80\pm0.14$&F\\
54927.6812&0.4757&$ -5.0$&$ -3.0$&            --  &           -- &B\\
55227.7110&0.7418&  19.0 &  20.0 &$ 16.39\pm0.21$&$-31.08\pm0.30$&B\\
55242.7476&0.7552&  18.0 &  18.8 &$ 17.95\pm0.30$&$-32.48\pm0.26$&B\\
55244.7298&0.7569&  15.0 &  18.0 &$ 17.84\pm0.37$&$-31.57\pm0.42$&B\\
55265.7237&0.7755&  22.0 &  22.0 &$ 21.03\pm0.23$&$-35.31\pm0.21$&B\\
55291.5929&0.7985&  26.0 &  27.0 &$ 21.77\pm0.23$&$-39.16\pm0.18$&B\\
55299.6685&0.8056&  23.0 &  23.0 &$ 22.54\pm0.25$&$-38.81\pm0.22$&B\\
55323.5905&0.8269&  25.9 &  26.3 &$ 24.80\pm0.12$&$-41.71\pm0.14$&F\\
55325.5834&0.8286&  23.6 &  24.0 &$ 24.26\pm0.10$&$-43.29\pm0.12$&F\\
55326.5139&0.8295&  25.3 &  25.7 &$ 24.49\pm0.10$&$-43.12\pm0.12$&F\\
55604.8774&0.0763&$-18.0$&$-17.9$&           --  &          --   &F\\
55641.6959&0.1090&$-22.9$&$-23.7$&$-23.24\pm0.12$&$ 41.69\pm0.18$&F\\
55643.6970&0.1108&$-26.5$&$-27.0$&$-24.54\pm0.11$&$ 41.74\pm0.13$&F\\
55698.6435&0.1595&$-23.6$&$-23.4$&$-24.28\pm0.13$&$ 42.72\pm0.17$&F\\
56068.5713&0.4876&$ -1.9$&$ -1.9$&$ -3.02\pm0.11$&$  4.82\pm0.15$&F\\
\hline\noalign{\smallskip}
\end{tabular}
\end{table*}

\begin{figure}
\resizebox{\hsize}{!}{\includegraphics{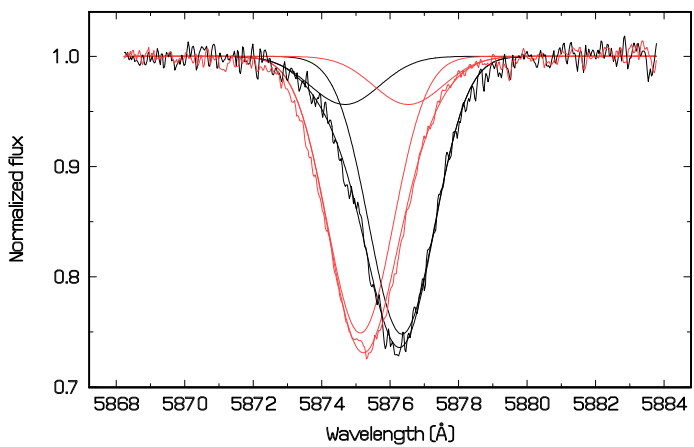}}
\caption{The primary and secondary components of HD~93146~A in the line
\ion{He}{i}~5876~\AA\ in spectra from RJD~54248 (black; phase 0.873) and 55643 (red; phase 0.111).}
\label{PriSec}
\end{figure}

\begin{table}
\begin{center}
\caption{Orbital elements for HD~93146~A; same notation as in Table~3.}
\label{orb93146}
\begin{tabular}{lr}
\hline\hline\noalign{\smallskip}
Element                             & Value            \\
\hline\noalign{\smallskip}
$P$ [days]                          &  1131.4\,(2.9)   \\
$T_{\rm periastr.}$ [RJD]           & 54385.9\,(6.6)   \\
$e$                                 & 0.4069\,(87)     \\
$\omega$ [$^\circ$]                 &  83.6\,(1.7)     \\
$K_1$ [\ks]                         &  25.73\,(37)     \\
$K_2$ [\ks]                         &  54.10\,(1.05)   \\
$\gamma_1$ [\ks]                    &   $-0.41\,(0.41)$\\
$\gamma_2$ [\ks]                    &   $+0.35\,(0.70)$\\
$m_1 \sin^3 i$ [\ms]                &    30.81\,(96)   \\
$m_2 \sin^3 i$ [\ms]                &    14.66\,(48)   \\
$a\,\sin i$ [\rs]                   &  1631\,(31)      \\
\vra [\ks]                          &  62\,(4)         \\
\vrb [\ks]                          &  50\,(10)        \\
rms$_1$ [\ks]                       &     0.91         \\
rms$_2$ [\ks]                       &     0.61         \\
\hline\noalign{\smallskip}
\end{tabular}
\end{center}

The spectra disentangled by \korel have RVs corresponding to the
systemic velocity. When used as templates for {\tt asTODCOR},
RVs corresponding to zero systemic velocity are obtained.
\end{table}

\begin{figure}
\resizebox{\hsize}{!}{\includegraphics{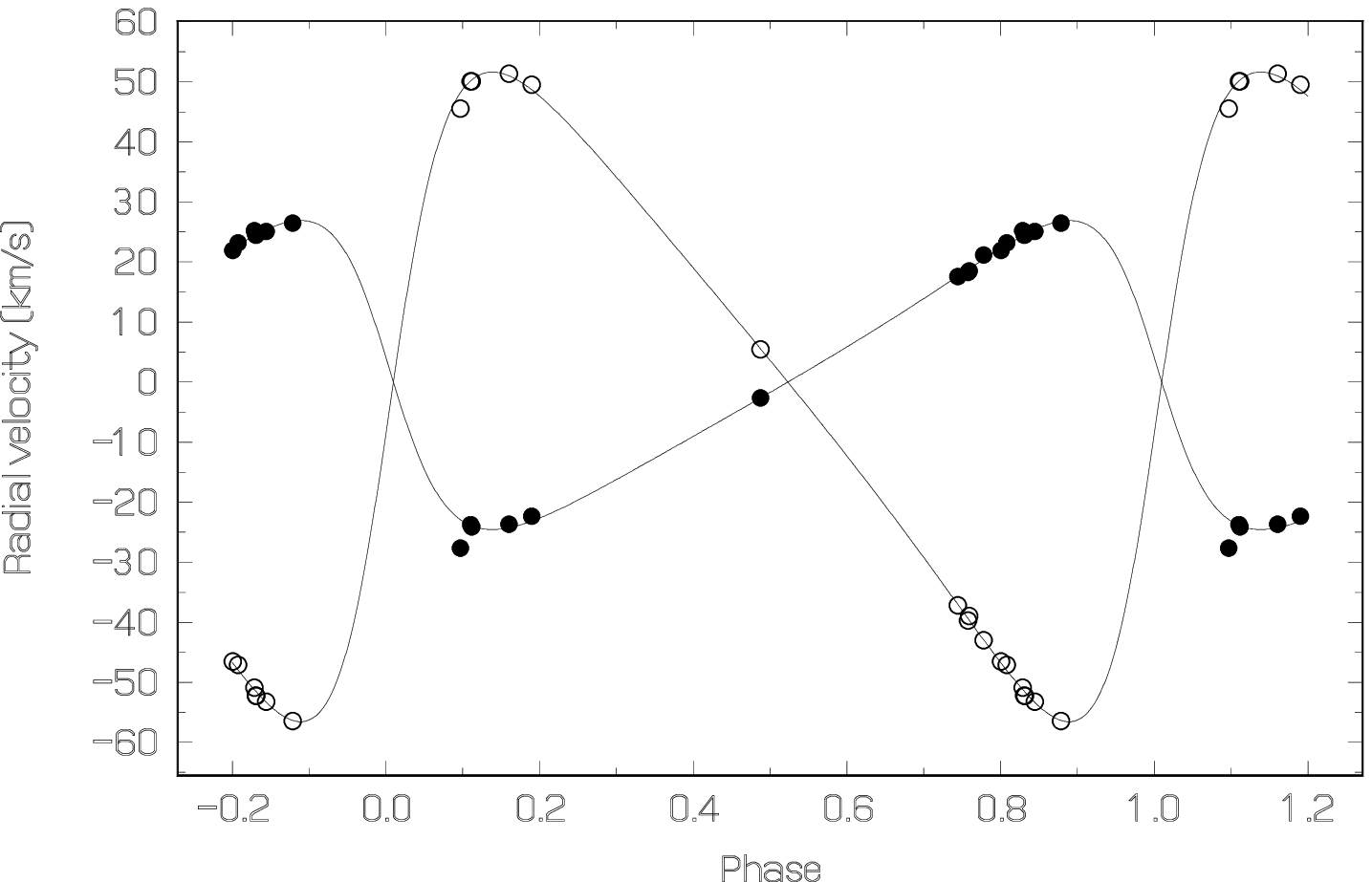}}
\caption{The orbital RV curves of HD~93146 based on {\tt asTODCOR} RVs.
The estimated errors are generally comparable to the size of the symbols
(black dots and open circles for the primary and secondary, respectively), and
are not, therefore, shown.}
\label{RV_93146}
\end{figure}

%\newpage

\section{HD~112244}

HD~112244 (HD 112244A, HR 4908, CD$-56^\circ4699$, HIP~63117 and 54~G~Cru) has
the spectral type O8.5\,Iab(f)p \citep{sota2014} and a visual brightness of
$V = 5$\m38. It has been part of the monitoring program for stellar multiplicity
\citep{chini} and attracted our attention because of its highly variable spectral
line profiles.

Radial velocities were measured by several investigators but no periodicity of RV
changes was reported. \citet{stick2001} measured RVs in 9 IUE spectra and tried to
search for periodicity in them and in previously published RVs, but could not find
any. They concluded that the observed RV changes are most likely due to atmospheric
instabilities. \citet{sota2014} in their GOSSS survey mention that according to
a parallel high-dispersion OWN survey \citep{barba2010}, \hd is a double-lined
spectroscopic binary with a 7\fd5 period. They remarked that they do not see
double lines in the GOSSS spectra.

\citet{jakate}, in his search for possible new $\beta$~Cep variables, found HD~112244
constant on six nights of his observations. However, the range of variations found
by the Hipparcos satellite \hp\ photometry from 5\m33 to 5\m41 \citep{esa97} seems to
indicate some micro-variability. Indeed, \citet{mar98} analysed \hp\ photometry and
suspected the presence of three possible periods, 4\fd518, 3\fd286 or 1\fd156, the
amplitude of variations being slightly over 0\m02 in all three cases. Repeated
analysis of the same data by \citet{koen2002} resulted in finding a period
of 2\fd00288 with a semi-amplitude of 0\m0085.

\begin{figure*}
\centering
\includegraphics[scale=0.32]{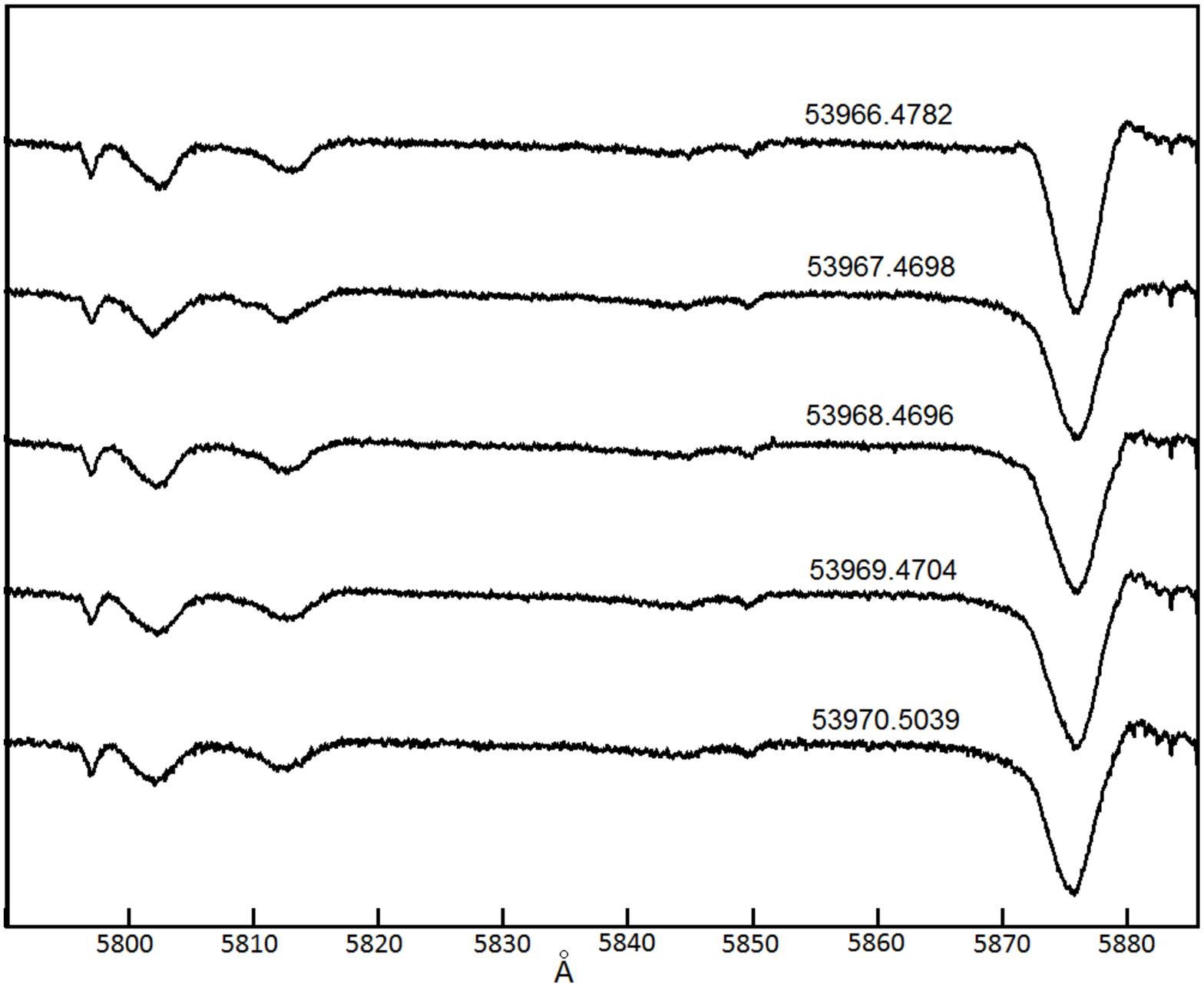}
\includegraphics[scale=0.32]{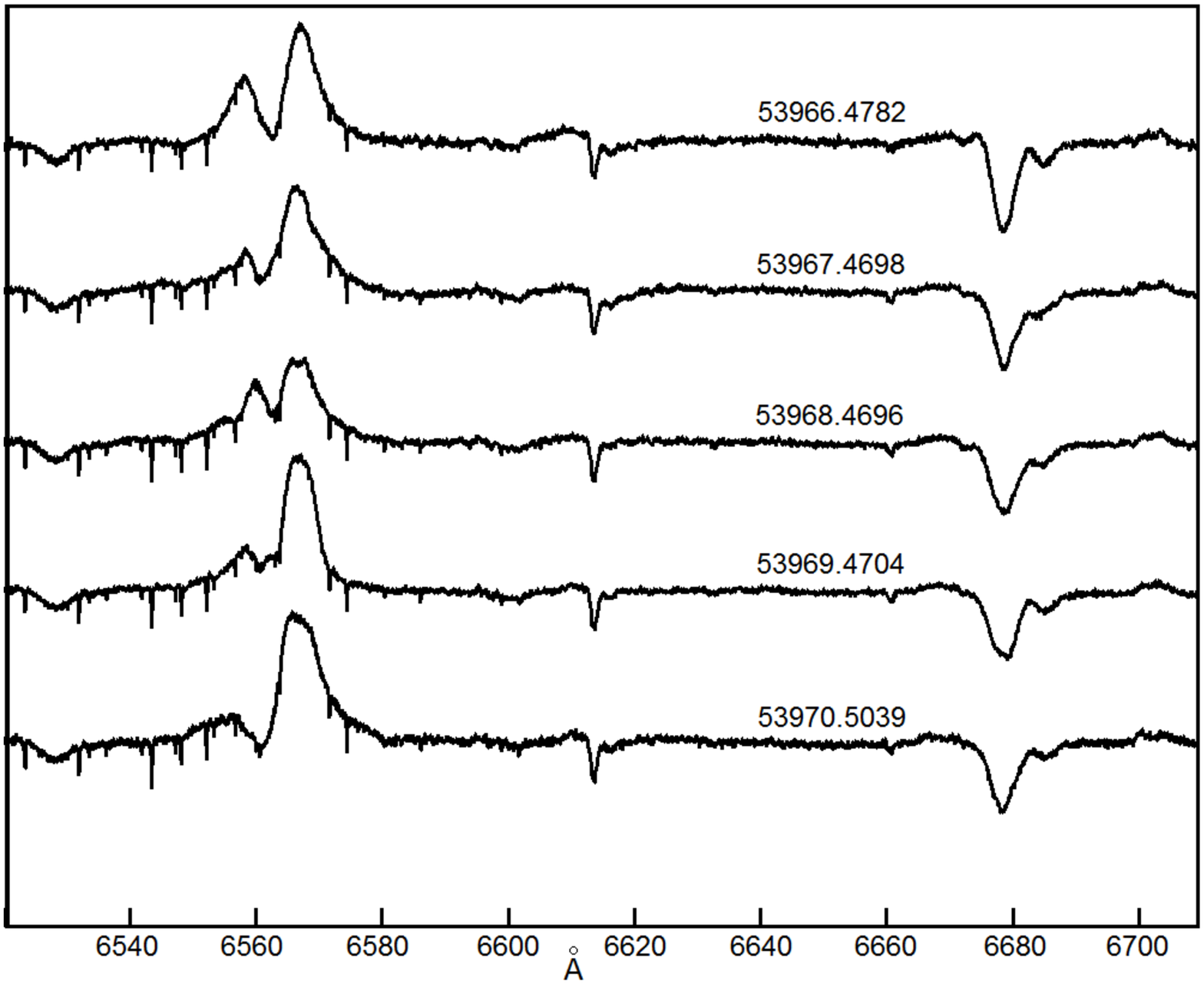}
\includegraphics[scale=0.32]{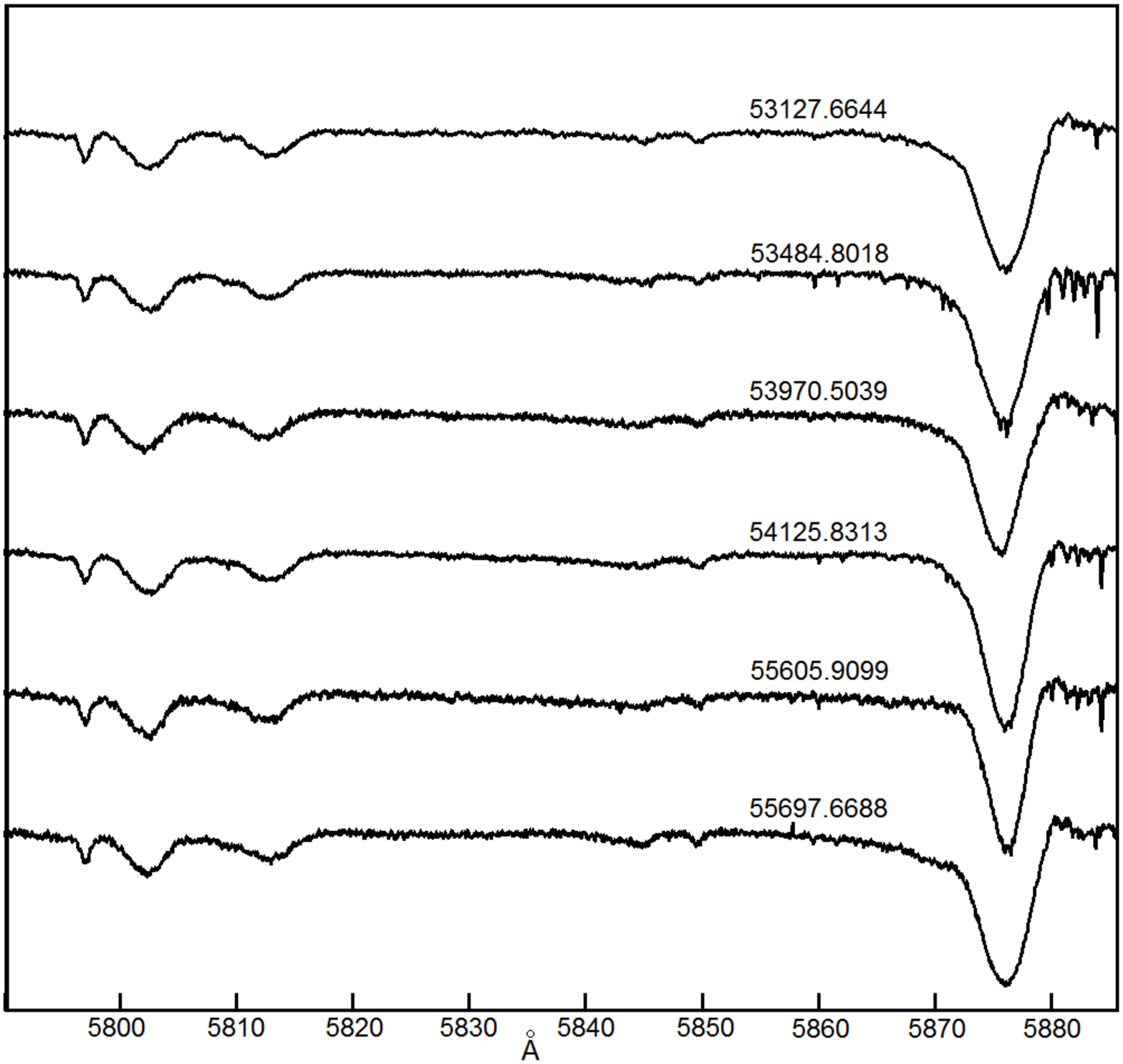}
\includegraphics[scale=0.32]{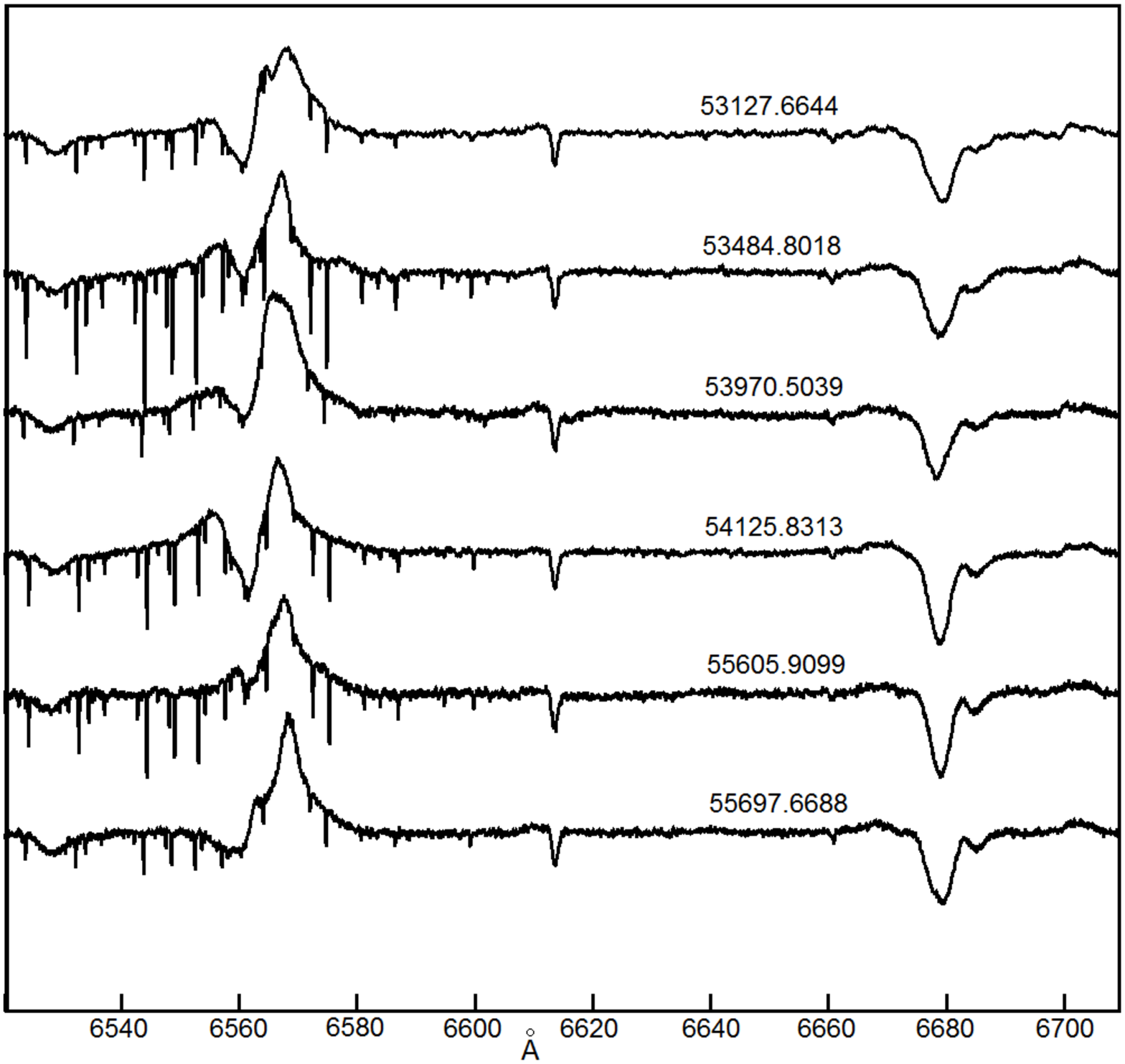}
\caption{Examples of variability of line profiles originating in the circumstellar
matter in the yellow and red parts of the spectra of \hde. Left panels show the
diffuse interstellar band at 5797.0~\AA, the \ion{C}{IV}~5801.3 and 5812.0~\AA,
and the \ion{He}{i}~5875.7~\AA\ line profiles, while the right panels show \hae,
the diffuse interstellar 6613.6~\AA\ band and the \ion{He}{i}~6678.2~\AA\ and
\ion{He}{ii}~6683~\AA\ line profiles for the same spectra. The upper panels
illustrate the variability on a time scale of days and the bottom panels show
variations on longer time scales. The offset of individual profiles is 0.25 of the continuum level.}
\label{redvar}
\end{figure*}

Our observational material consists of 68 BESO and 30 FEROS spectra secured
between May 2004 and August 2014 (RJD~53127 - 56880). Ten FEROS spectra secured
between RJD~53898.6647 and 53919.5607 have exposure times
of 9~s only and are relatively noisy. Also, the last series of BESO spectra
starting with RJD~56854.4909 was obtained in very adverse weather conditions
and has a lower quality. For that reason, we measured the S/N for each spectrum
and applied weights proportional to the square of S/N and normalised to the mean S/N.

HD~112244 is an emission-line object and has strong stellar wind.
\citet{martins2014} have shown that rapid and apparently irregular variability of
the wind line profiles is characteristic of the OB supergiants. This is also true
for \hd as illustrated in Fig.~\ref{redvar}, where we show  parts of the yellow
and red spectra for selected spectrograms. It is seen that the strength and shape
of the \ha line and also of the faint emission at the red wing of the
\ion{He}{i}~5876~\AA\ line both vary on timescales from days to years, while the
line profiles of high-ionisation species as \ion{He}{ii} or \ion{C}{iv} remain
relatively stable. This fact represents a complication for RV measurements
since even seemingly photospheric lines can be partly affected by these physical variations.

To check whether \hd is indeed a spectroscopic binary, we measured RVs of several
spectral lines using different techniques. The RVs of basically photospheric
lines of \ion{He}{i}~4921.9~\AA\ and 6678.2~\AA, \ion{He}{ii}~5411.5~\AA,
\ion{C}{iv}~5801.3~\AA\ and 5812.0~\AA, and of two lines affected by emission
(\ha and \ion{He}{i}~5875.7~\AA) were measured via a comparison of direct and
flipped line profiles in \spefoe. The RV of the clean \ion{He}{ii}~5411.5~\AA\
line was also measured through
a~Gaussian fit of the profile. Using \citet{stw78} phase dispersion minimisation
period search technique, we analysed all these RV data subsets for periodicity
over the frequency range down to 10~c\,d$^{-1}$. In all cases, this led to the
detection of a dominant period of 27\fd67 and RV curves indicative of a highly
eccentric orbit.

For the final analysis, we selected the region from 5792 to 5855~\AA\ and
proceeded in the following way. We formally disentangled the stellar spectrum
of the \ion{C}{iv}~5801 and 5812~\AA\ and the spectrum of diffuse interstellar
lines. These disentangled spectra, shown in Fig.~\ref{civ-is}, were then used
as templates for the 2D cross-correlation with the program {\tt asTODCOR}.

\begin{figure}
\centering
\resizebox{\hsize}{!}{\includegraphics{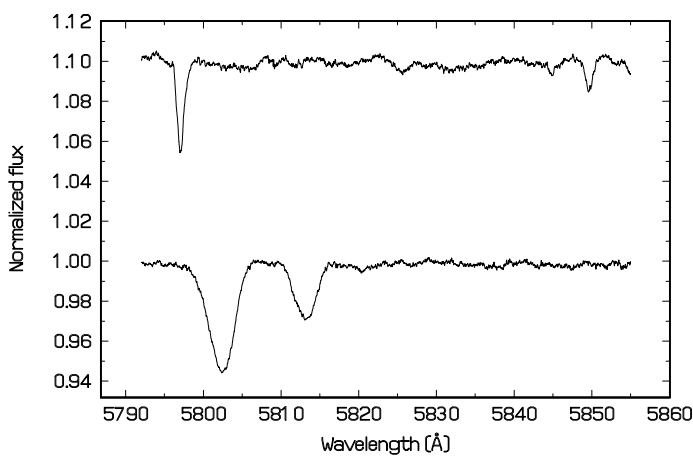}}
\caption{The disentangled spectrum of HD~112244 and of diffuse interstellar
 bands in the wavelength region 5792 -- 5855~\AA.}
\label{civ-is}
\end{figure}

\begin{figure}
\centering
\resizebox{\hsize}{!}{\includegraphics{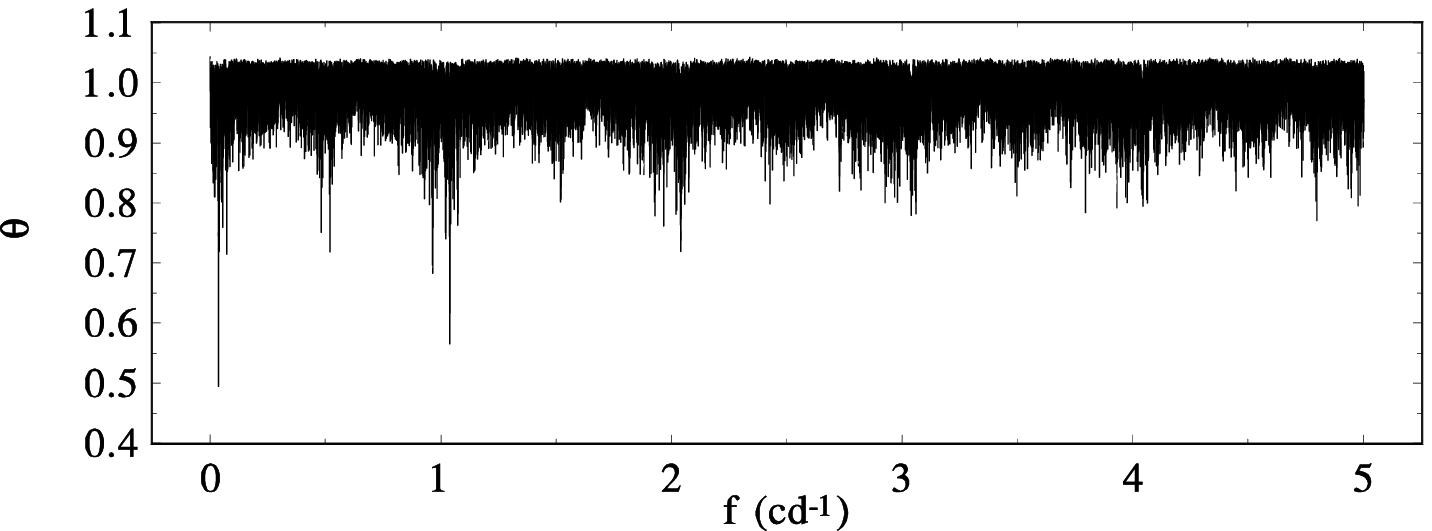}}
\resizebox{\hsize}{!}{\includegraphics{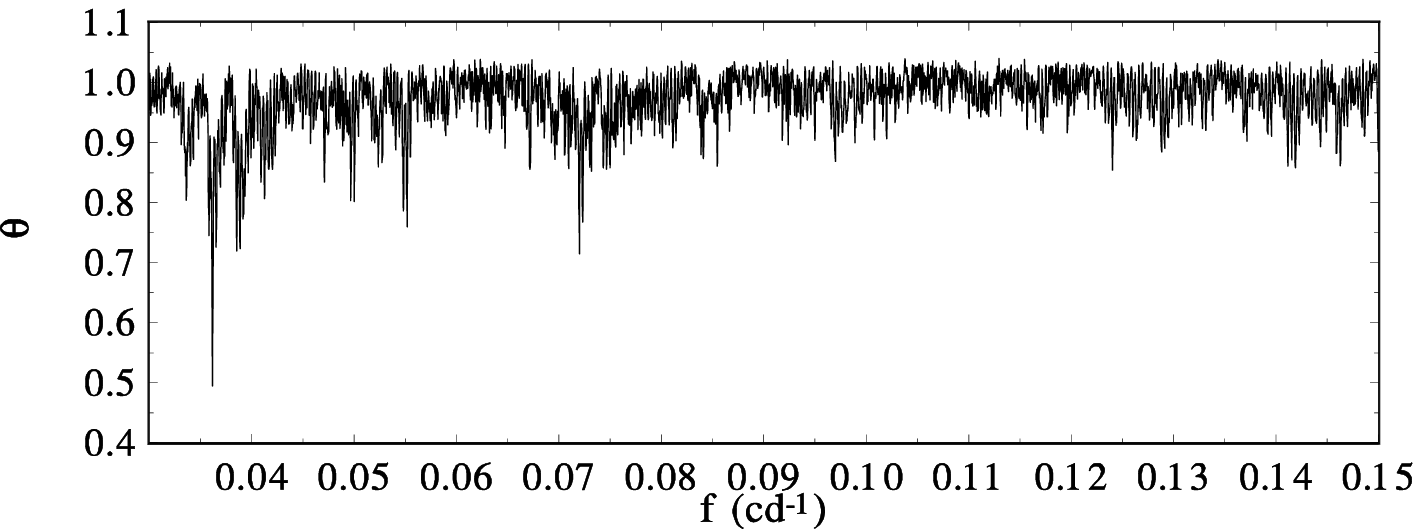}}
\caption{Stellingwerf's $\theta$ statistics periodogram for {\tt asTODCOR}
RVs from the 5792 -- 5855~\AA\ region. A complete periodogram for periods down
to 0\fd2 is shown in the upper panel and a zoomed part of it in the bottom one.
}
\label{the-civ}
\end{figure}

Figure~\ref{the-civ} shows the {phase-dispersion} periodogram for
{\tt asTODCOR} RVs down to a period of 0\fd2. One can see that the period of
27\fd67 (frequency 0.036~c\,d$^{-1}$) represents the dominant minimum in
the periodogram. Other deeper minima can be identified with the second
harmonic at 0.072~c\,d$^{-1}$, 1~d alias at 0.963~c\,d$^{-1}$ and 1~yr alias
at 0.039~c\,d$^{-1}$. We note that there is no evidence of the earlier
reported 7\fd5 period ($f=0.133$~c\,d$^{-1}$). Our orbital solution based
on these {\tt asTODCOR} RVs with weights proportional to the square of
the S/N is summarized in Table~\ref{orb112244}.

\begin{table}
\begin{center}
\caption{Orbital elements for HD~112244; same notation as in Table~3.}
\label{orb112244}
\begin{tabular}{lr}
\hline\hline\noalign{\smallskip}
Element                             & Value          \\
\hline\noalign{\smallskip}
$P$ [days]                          &  27.6652(22) \\
$T_{\rm periastr.}$ [RJD]           & 55532.20(21)   \\
$e$                                 & 0.692\,(50)    \\
$\omega$ [$^\circ$]                 &  290.8\,(7.5)  \\
$K_1$ [\ks]                         &  17.4\,(1.9)   \\
$f(m)$ [\ms]                        &   0.00572\,(22)\\
\vra   [\ks]                        & 122 -- 166$^1$ \\
rms   [\ks]                            & 5.8\\
\hline\noalign{\smallskip}
\end{tabular}
\end{center}

$^1$ The interval corresponds to full width at half maximum (FWHM).
\end{table}

\noindent
The secondary must be much less massive and indeed our attempts to disentangle
it failed. The corresponding RV curve is depicted in Fig.~\ref{civ}. It is
clear that larger deviations from the curve come invariably from spectra of
lower quality. Regrettably, no RVs of superb quality are available at
the bottom part of the RV curve. We postpone the discussion of binary
properties to the future, when a dedicated series of well exposed spectra will
be obtained. Nevertheless, a crude estimate of the secondary mass can be
obtained from the mass function $f(m)$. Adopting
$m_1 = 31.54$ for O8.5\,I \citep{martins}, we obtain a mass range of
$1.85 - 3.85$\,\ms\ for a range of inclinations from $90^\circ$ to $30^\circ$.

\begin{figure}
\centering
\resizebox{\hsize}{!}{\includegraphics{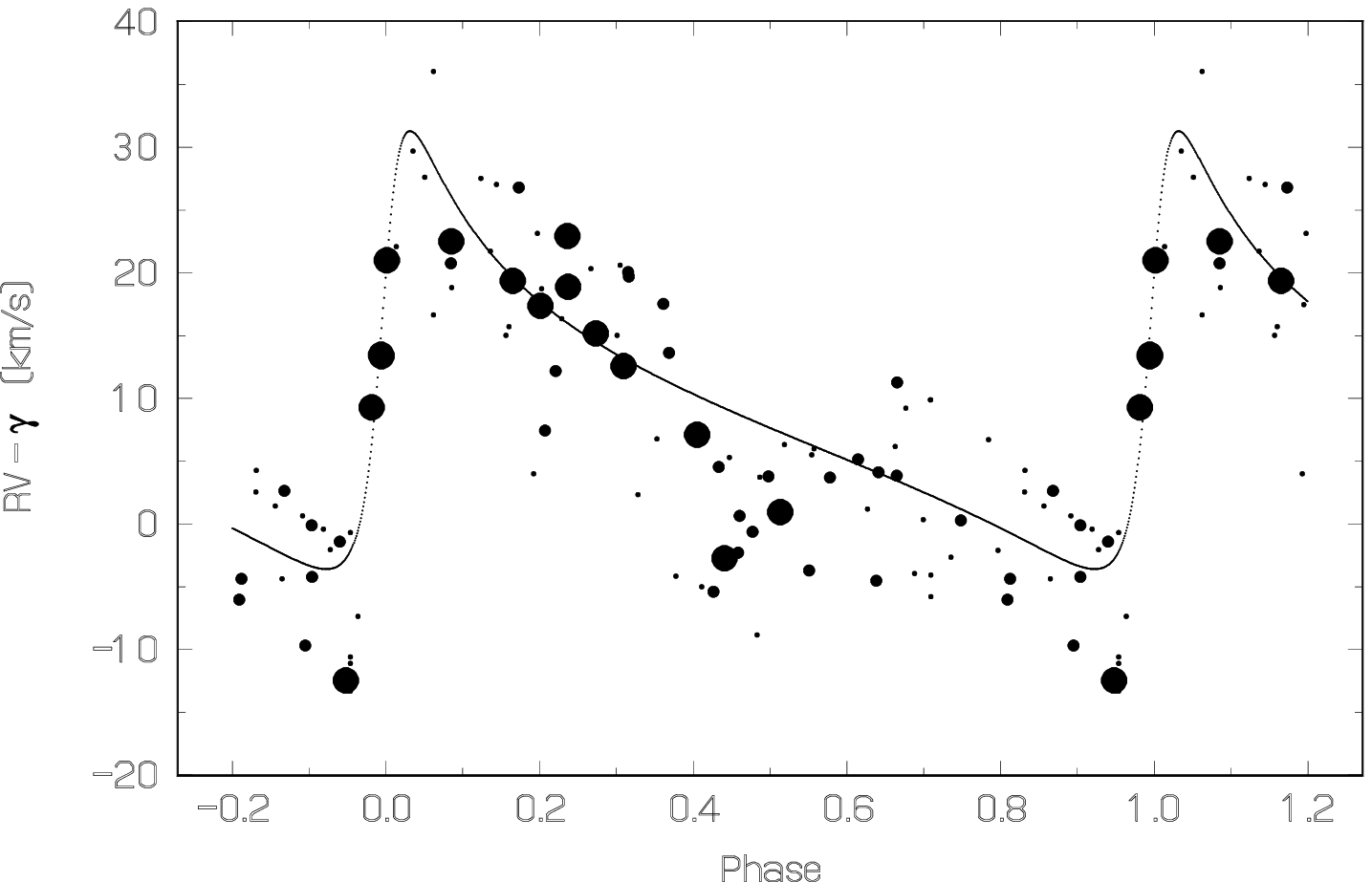}}
\resizebox{\hsize}{!}{\includegraphics{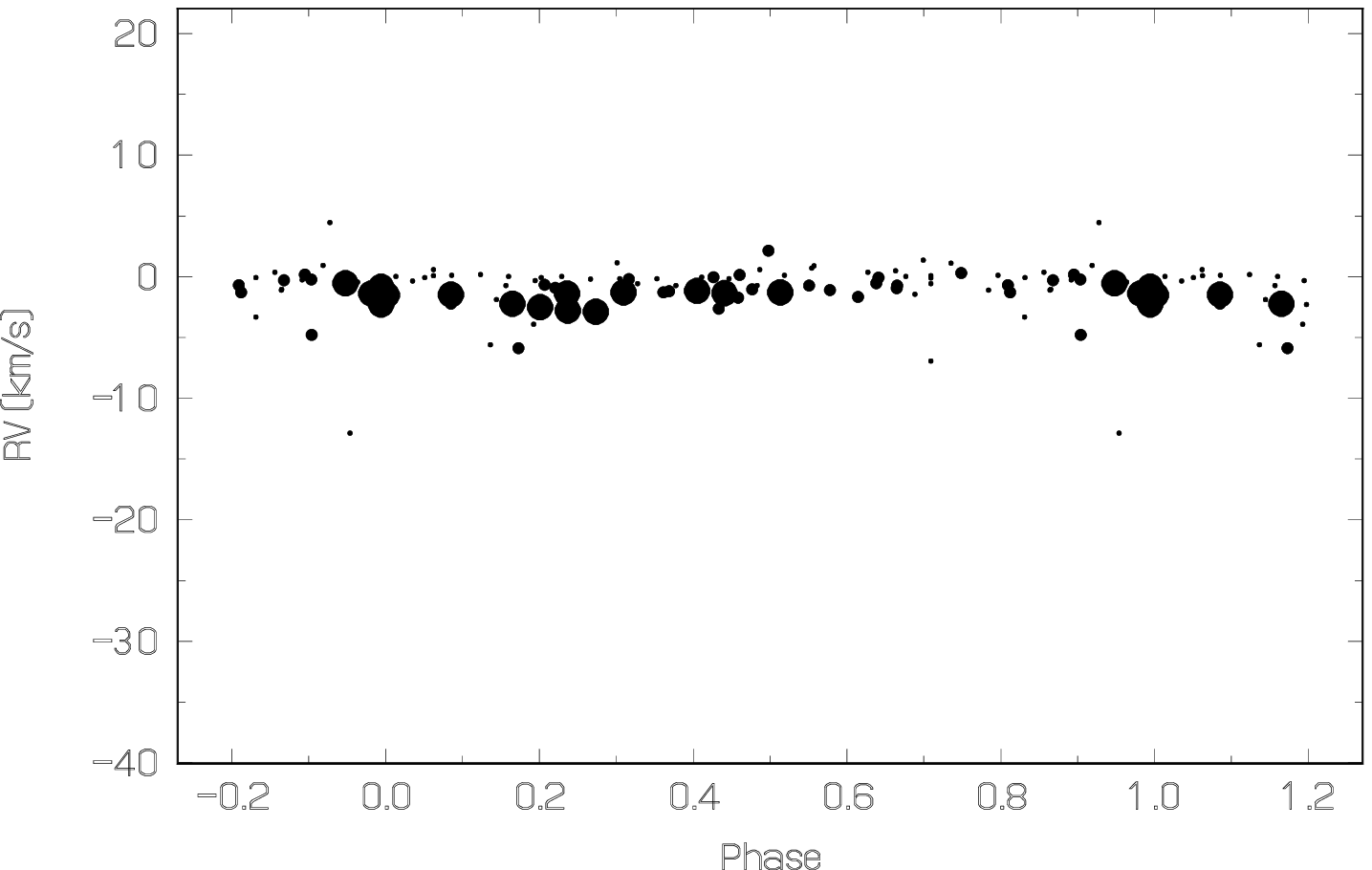}}
\resizebox{\hsize}{!}{\includegraphics{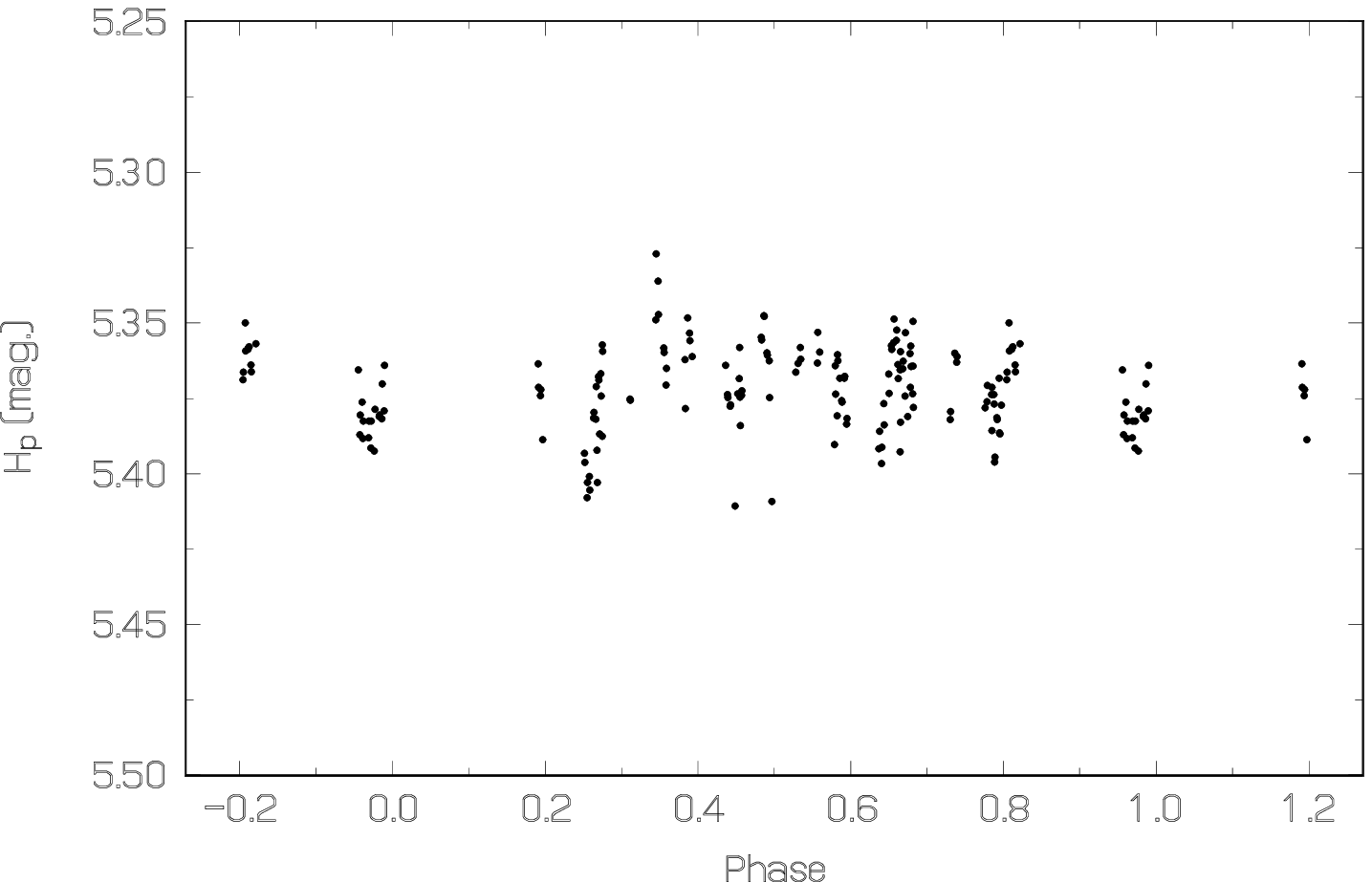}}
\caption{{\sl Upper panel:} the orbital RV curve plot of HD~112244 based on
{\tt asTODCOR} RVs from the 5792 -- 8555~\AA\ region. RVs from spectra with
weights higher than 1.5 are shown by large black circles, those with weights
between 1 and 1.5 with smaller black circles, and those with weights lower
than 1 with black dots.
{\sl Middle panel:} {\tt asTODCOR} RVs of the diffuse absorption bands
shown on the same scale to characterise the uncertainties of the RV
determination. The same symbols as above.
{\sl Bottom panel:} Hipparcos $H_{\rm p}$ magnitude plotted vs. orbital phase.}
\label{civ}
\end{figure}

The periodic RV variability of the stellar lines seems to be beyond doubt,
especially if compared to the RVs of the diffuse interstellar bands.
The light variations plotted vs. orbital phase appear to be indicative of some
 co-rotating structures, also known for some other OB stars
\citep[cf., e.g.][]{owocki95,hec89,hec99}. More observations are required, however,
to prove or disprove this classification.

\section{HD~123056}

HD~123056 (CPD$ -59\degr 5404$, CD$ -59\degr 5090$, LS~3200) is a field star with
$V = 8.14$ and was classified as O9.5\,V((n)) \citep{Walborn73}; \citet{sota2014}
give an integral spectral type of O9.5\,IV(n). \citet{feast} published four RVs
and noted their variability; \citet{chini} found varying double lines and
assigned the status SB2. \citet{sota2014} suggested that the star is ``at least SB2'',
speculating that it might also be a triple or multiple system.

We could secure ten spectra with BESO; five more spectra were obtained within
 the program "Tycho Brahe" in May 2015, when the 2.2~m MPI telescope was
allocated to programs proposed by Czech astronomers (see Table~\ref{123}).
Furthermore, there are eight FEROS spectra in the ESO archive.
 The profiles of the \ion{He}{i}~6678~\AA\
and the \ion{He}{ii}~5411~\AA\ lines (FEROS spectra) are shown in Fig.~\ref{F123}.

\begin{figure}[h!]
\resizebox{\hsize}{!}{\includegraphics{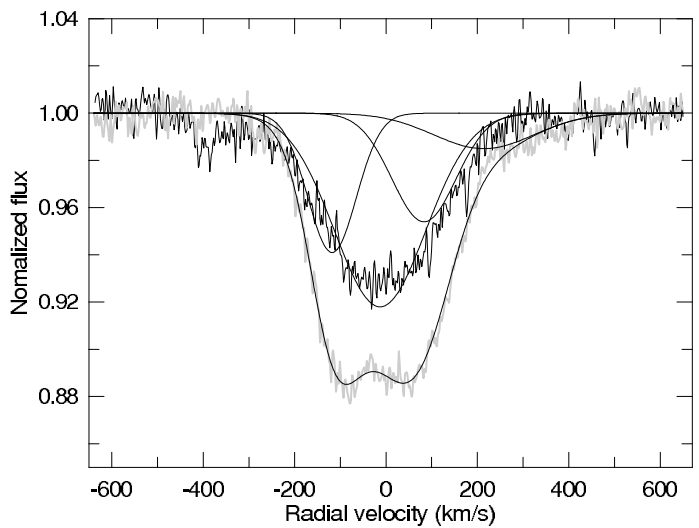}}
\caption{An attempt to resolve the profile of the line \ion{He}{i}~6678~\AA\ in
 HD~123056 obtained at RJD~54960. The RV of the central line of the profile
(binary A) has been assumed to be identical to the RV measured for the line
 \ion{He}{ii}~5411~\AA. The thick line shows the \ion{He}{ii} line 5411
(with DIBs at approximately $+390$ and $-390$ \ks). The grey line is the
 observed line \ion{He}{i}~6678~\AA\ and the thin lines represent components
of the binary B. The line \ion{He}{ii}~6683~\AA\ in the binary A is also present.}
\label{F123}
\end{figure}

As the \ion{He}{ii}~5411~\AA\ line is broad and any indications for duplicity are
 virtually missing, we conclude that it belongs to a third component, hereafter
referred to as ``A''. From its position changes, we conclude that A has a~variable RV.
The number of spectra is not sufficient for a reliable determination of the period.
With the Radcliffe observatory spectra included, the period-finding program found
 a period near 1310 days as the best one and the free orbital solution improved this
 value to $1314.1\pm1.8$\,d; {the probable orbital elements are listed in Table~8}.
We admit that this must not be the only possible solution, since some of the RVs are
 not completely reliable. Components of the binary~B are distinguishable near
 elongations of their orbit in the \ion{He}{i} lines (cf Fig.~\ref{F123}).

\begin{table*}
\caption{\spefo RVs of the wings and two line cores of
the \ion{He}{i}~5876~\AA\ line and RVs from Gaussian fits of
the \ion{He}{ii}~5411~\AA\ line of HD~123056 (in \ks).
The four early Radcliffe RVs (R) are the mean RVs of several spectral lines.}
\label{123}
\begin{tabular}{crrrrrrc}
\hline\hline\noalign{\smallskip}
RJD                & 5876   &  \oc  & 5876   &  5876       & 5411 & \oc  & Spectro-\\
                & wings  &  core & 2$^{\rm nd}$ core& & &       & graph   \\
\hline\noalign{\smallskip}
34825.5256          &$-30.00$&$ -5.0$&& &&        & R       \\
34834.4991          &$-23.00$&   3.3 && &&        & R       \\
35181.5451          &$-59.00$&$ -8.3$&& &&        & R       \\
35208.4685          &$-44.00$&   5.0 && &&        & R       \\
54935.6907          &$-38.23$&   9.7 &$ -2.10$& --     &$ -36 $ &   9.8 & B       \\
54952.7769          &$-53.41$&$ -6.8$&$ -3.59$& --     &$ -48 $ &$ -3.3$& B       \\
54954.7400          &$-52.74$&$ -6.2$&  19.43 &$-88.48$&$ -53 $ &$ -8.4$& F       \\
54960.7492          &$-48.83$&$ -2.8$&$-47.60$& --     &$ -51 $ &$ -6.8$& B       \\
54976.6593          &$-39.31$&   5.5 &$-49.72$& --     &$ -38 $ &   5.1 & F       \\
55240.9051          &$-10.76$&  12.1 &  28.73 &$-68.46$&$ -12 $ &  10.9 & B       \\
55241.9183          &$-18.94$&   3.8 &  14.42 & --     &$ -23 $ &$ -0.2$& B       \\
55251.8066          &$-19.02$&   3.0 &$-31.34$& --     &$ -22 $ &   0.1 & B       \\
55253.8783          &$-35.55$&$-13.7$&  19.24 &$-96.36$&$ -28 $ &$ -6.1$& B       \\
55605.8938          &$-11.34$&$ -4.6$&   7.84 &$-89.40$&$ -13 $ &$ -7.8$& F       \\
55642.7878          &$-14.63$&$ -7.3$&$ -3.83$& --     &$ -11 $ &$ -5.2$& F       \\
55697.7150          &$ -1.42$&   8.0 &$-60.12$&  22.53 &$  -2 $ &   6.2 & F       \\
55698.7538          &$ -5.48$&   4.0 &$ -4.33$& --     &$  -5 $ &   3.2 & F       \\
55794.9821          &$-16.54$&   1.0 &   3.44 & --     &$ -15 $ &   2.8 & B       \\
55803.9761          &$-13.59$&   5.0 &$-37.69$& --     &$ -16 $ &   3.1 & B       \\
55805.9746          &$-12.87$&   6.0 &$  7.33 $& --     &$  -7 $ &  12.3 & B       \\
56067.7858          &$-48.14$&   4.2 &$-57.94$& --     &$ -45 $ &   5.5 & F       \\
56068.7840          &$-52.54$&$ -0.1$&$ -4.48$& --     &$ -53 $ &$ -2.5$& F       \\
57155.8197          &$-24.22$&$ -0.6$&$-41.14$&  29.73 &$ -33 $ &$ -8.3$& F       \\
57156.8305          &$-27.45$&$ -3.7$&$-38.31$& --     &$ -28 $ &$ -3.2$& F       \\
57157.7282          &$-19.56$&   4.4 &  24.60 &$-88.29 $&$ -27 $ &$ -2.0$& F       \\
57159.7152          &$-26.52$&$ -2.3$ &  18.48 & --     &$ -26 $ &$ -0.7$& F       \\
57160.5437          &$-33.10$&$ -8.8$&$-56.91$& --     &$ -30 $ &$ -4.6$& F       \\
\hline\noalign{\smallskip}
\end{tabular}

To estimate the measurement errors of the components of
\ion{He}{i}~5876~\AA\ in SPEFO, we repeated the measurements several
times independently. The results for the wings usually differed by less
than 1~\ks, while the settings for the line cores are less certain,
especially for lower S/N BESO spectra.
\end{table*}

It is not clear whether this long period is that of a binary or of
the third body orbit (A) around the closer binary (B). Nevertheless,
the body with the long period must be of earlier spectral type
(say $\approx$ O8.5). As the components of the binary B are better seen
in the \ion{He}{i} lines (are invisible in \ion{He}{ii}~5411~\AA),
they are of type $\approx$ B0; a discrimination between components
B1 and B2 is difficult. The amplitudes of RVs in all orbits are rather small,
therefore the accuracy of the derived periods is also affected.
The period of the binary B seems to be slightly shorter than 2 days.
To summarise, HD~123056 appears to be another hierarchical triple system
of high-mass stars.

\begin{table}
\begin{center}
\caption{Probable orbital elements of the long AB orbit of HD~123056;
same notation as in Table 3.}
\label{orb1230}
%\begin{minipage}
\begin{tabular}{lr}
\hline\hline\noalign{\smallskip}
Element                             &  Value             \\
\hline\noalign{\smallskip}
$P$ [days]                          &  1314.1 (1.8)      \\
$T_{\rm periastr.}$ [RJD]           & 54648\,(95)        \\
$e$                                 &     0.182\,(70)    \\
$K_1$ [\ks]                         &    23.4\,(2.7)     \\
$\gamma_1$ [\ks]                    &  $-28.0\,(1.5)$    \\
$f(m)$ [\ms]                        &    1.66(72) \\
$a_1\,\sin i$ [\rs]                 &   598(76)       \\
\vra [\ks]                          &  140\,(20)         \\
\vrb [\ks]                          &  140\,(20)         \\
\vrc [\ks]                          &  225\,(30)         \\
rms$_1$  [\ks]                      &    5.85            \\
\hline\noalign{\smallskip}
\end{tabular}
\end{center}
\end{table}

\begin{figure}
\resizebox{\hsize}{!}{\includegraphics{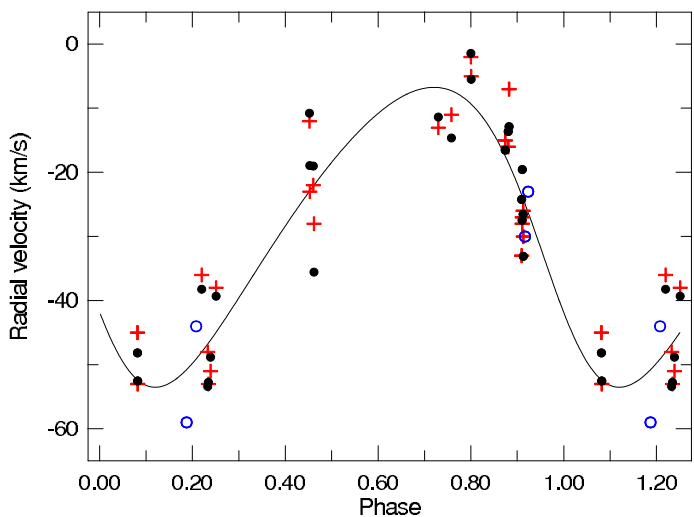}}
\caption{The long orbit of HD~123056 (binary A). The curve corresponds to
 elements $P=1314 $ days, $e=0.18$, $K=23$~\ks, $V_\gamma=-27$~\ks). Black
full circles  indicate measurement of the line \ion{He}{i} 5876~\AA\ while
 red plus symbols represent  measurement of the line \ion{He}{ii} 5411~\AA\
and blue open circles  indicate Radcliffe data.}
\label{rv_123}
\end{figure}

\section{HD~164438}

HD~164438 (BD$ -19\degr 4800$, LS~4567) is a star in the Sgr~OB1 association
with $V = 7.48$ and a spectral type of O9.2\,IV \citep{sota2014}.
\citet{mason98} found a possible binary companion at a separation of 0\farcs05,
which, however, needs confirmation as it was not seen by \citet{sana2014}.
\citet{chini} classified HD~164438 as SB1, which was recently corroborated
by OWN data \citep{sota2014}.

We were able to secure 15 BESO spectra and downloaded 5 FEROS spectra from
the ESO archive. The RVs were measured from the \ion{He}{i}~5876~\AA\ and
\ion{He}{ii}~5411~\AA\ lines via Gaussian fits. The corresponding results
are listed in Table~\ref{RV_164}, the orbital elements are given in
Table~\ref{orb164438}, while the RV curve is shown in Fig~\ref{RV164}.
When inspecting the line profiles, one cannot find any trace of the lines
 from the secondary. We attempted to detect the secondary through
disentangling with \korele, but without success. This means that the EWs of
 the secondary component must be approximately $ 20$ times smaller than
those of the primary, implying a spectral type later than B3. Using the
mass function $f(m)$ and adopting $m_1 = 19$\,\ms \ for O9.2\,IV
{\citep[as can be obtained by intepolating in tables by, e.g. ][]{martins}}
 we obtain a mass range of $2.02 - 4.32$\,\ms\ for a range of inclinations
 from $90^\circ - 30^\circ$.

\begin{table}
\caption{RVs based on the Gaussian fits of \ion{He}{i}~5876~\AA\ line
of the primary component of HD~164438 (in \ks).}
\label{RV_164}
\begin{tabular}{clrrc}
\hline\hline\noalign{\smallskip}
RJD        & Phase  & RV with error & \oc & Spg.\\
\hline\noalign{\smallskip}
54600.8993 & 0.1841 & $  6.5\pm1.3$&  0.6 & F       \\
54937.8665 & 0.0576 & $-19.0\pm2.2$&  0.6 & B       \\
54943.8670 & 0.6430 & $ -2.7\pm1.8$&  1.5 & B       \\
54976.9109 & 0.8667 & $-33.6\pm1.6$&  0.8 & F       \\
55072.6480 & 0.2065 & $  6.6\pm1.8$&$-1.7$& B       \\
55121.5368 & 0.9759 & $-39.7\pm1.8$&$-0.1$& B       \\
55697.9411 & 0.2082 & $  8.0\pm1.2$&$-0.8$& F       \\
55698.8548 & 0.2973 & $ 13.1\pm1.2$&  0.7 & F       \\
55798.6818 & 0.0362 & $-27.8\pm2.0$&$-1.4$& B       \\
55812.6264 & 0.3966 & $ 10.2\pm1.8$&$-0.9$& B       \\
55813.6076 & 0.4923 & $  6.4\pm1.9$&$-0.4$& B       \\
55814.6200 & 0.5910 & $ -1.0\pm1.9$&$-1.0$& B       \\
55815.6204 & 0.6886 & $ -9.7\pm1.6$&$-0.3$& B       \\
55816.6211 & 0.7863 & $-23.3\pm1.8$&$-1.4$& B       \\
55817.6199 & 0.8837 & $-37.3\pm1.6$&$-0.9$& B       \\
55818.6224 & 0.9815 & $-38.3\pm1.6$&  0.6 & B       \\
55819.6209 & 0.0789 & $-13.0\pm1.6$&  0.8 & B       \\
55821.6047 & 0.2724 & $ 11.9\pm1.7$&$-0.1$& B       \\
56067.9290 & 0.3031 & $ 13.6\pm1.3$&  2.0 & F       \\
56441.9326 & 0.7898 & $-21.3\pm1.7$&  1.2 & B       \\
\hline\noalign{\smallskip}
\end{tabular}
\end{table}

\begin{figure}
\resizebox{\hsize}{!}{\includegraphics{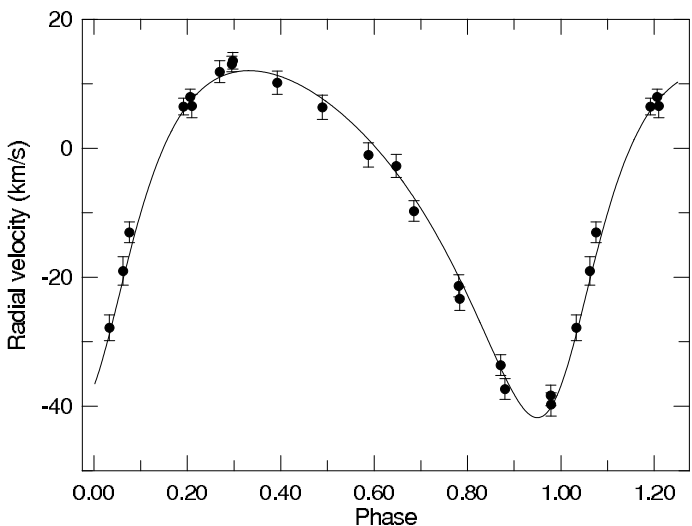}}
\caption{The RV curve of HD~164438 based on Gaussian fits of
the \ion{He}{i}~5876~\AA\ line plotted for the ephemeris of Table~\ref{orb164438}.}
\label{RV164}
\end{figure}

\begin{table}
\begin{center}
\caption{New orbital elements for HD~164438; same notation as in Table~\ref{orb92206}.}
\label{orb164438}
\begin{tabular}{lr}
\hline\hline\noalign{\smallskip}
Element                     & Value             \\
\hline\noalign{\smallskip}
$P$ [days]                  &    10.25042\,(40) \\
$T_{\rm periasrt.}$ [RJD]   & 55501.05\,(6)     \\
$e$                         &     0.310\,(13)   \\
$\omega$ [$^\circ$]         &   218.5\,(2.5)    \\
$K_1$ [\ks]                 &    26.9\,(4)      \\
$\gamma_1$ [\ks]           & $-8.0\,(3)$       \\
$f(m)$ [\ms]        & 0.0178\,(8) \\
\vra [\ks]          & 89\,(2)          \\
rms$_1$  [\ks]                  &     1.1           \\
\hline\noalign{\smallskip}
\end{tabular}
\end{center}
\end{table}

\section{HD~164816}

HD~164816 (CPD$ -24\degr 6146$, CD$ -24\degr 13816$, LS~4597; $V=7.08$) is
located towards the young open cluster NGC~6530. It was originally
 classified as O9.5\,III--IV(n) \citep{Walborn73}. The spectral type
B3\,Ve assigned by \citet{Levenhagen06} must be a mistake, since the
 O-type character of the spectrum is without any doubt; the emission
is present as narrow lines, so it originates from the surrounding H~II region.
\citet{chini} classified HD~164438 as SB2. Double lines are also seen by
\citet{sota2014}, who derived spectral types of O9.5\,V + B0\,V for the two components.

\citet{trepl} have published an extensive multi-wavelength (X-ray, $\gamma$-ray,
 optical and radio) study of the binary HD~164816. They find an orbital
 period of 3.82 days and projected masses of $m_1\sin^3i = 2.355(69)$\,\ms\,
 and $m_2\sin^3i = 2.103(62)$\,\ms. They also identified an associated X-ray
 source with a soft X-ray excess and a 10\,s pulsation, which they interpret
 as the signature of a neutron star in the system.

During an interferometric infrared survey, \citet{sana2014} detected
a~faint companion separated by 57\,mas. This object is not a component of
the SB2 system itself because this pair is separated by only 0.07\,mas
assuming a distance of 1\,kpc as suggested by \citet{trepl}. The
difference of the $H$ magnitude of HD~164816 and the faint companion
 is $\Delta H = 3.4$, suggesting quite different masses. Because the
 object is also detected in X-rays, \citet{sana2014} suggested that
they probably see an active later-type object, which may provide an
alternative explanation to the X-ray excess of HD~164816.

We re-analysed the existing BESO spectra and also included two FEROS spectra
 from the ESO archive in our analysis. In general, we can confirm the orbital
 elements published by \citet{trepl}. However, we obtain a systemic velocity
equal to the velocity of the cluster. We suspect that in the cited paper the
spectra were not reduced correctly. Note that \citet{stick2001} published
 the following RVs (in \ks)

\smallskip
RJD 43782.337: $V_1 = 39.7$, $V_2 = -34.5,$

RJD 44899.205: $V_1 = 77.7$, $V_2 = -81.9,$

\smallskip\noindent
which clearly suggest that the systemic velocity is close to zero (or close
to the RV of the cluster, which is approximately $-7$~\ks). We believe
that the reason why \citet{trepl} obtained an anomalous systemic velocity
lies in the fact that they overlooked that the first pixel of BESO spectra
is numbered $-51$ in their reduction and therefore the spectra must be
shifted by 1.56~\AA. For the \ion{He}{i}~5876~\AA\ line, this corresponds
 to 78~\ks, which is close to the value of the erroneous systemic velocity
 published by \citet{trepl}.

As a consequence, the distance of the binary might well be equal to the
distance of NGC\,6530 whose latest determinations are 1250\,pc
 \citep{Prisinzano05} and 1322\,pc \citep{Kharchenko05}, respectively.
 Likewise, the cluster was found to be affected by anomalous extinction
with $R = 4.5$ \citep{Fernandes12}. Although SIMBAD classifies HD~164816
 as a Be star \citep{Levenhagen06}, it seems that among eight sources of
$UBV$ photometry there are no pronounced light or colour changes observed.
We therefore used mean photometric values $V = 7.092$ and $B-V = 0.02$
 \citep{Mermilliod86}. De-reddening leads to $E_{B-V} = 0.292$,
 $(B-V)_0 = -0.29$ and $(U-B)_0 = -1.09$ resulting in colours that
agree well with an O9.5\,V + B0\,V spectrum. At a distance of 14\,arcmin
from the cluster centre and a cluster radius of 15\,arcmin, the star is
located at the outskirts of NGC\,6530. Using $R = 4.5$ to derive the
 visual extinction yields $V_0 = 5.78,$ while a normal value of
$R = 3.1$ would imply $V_0 = 6.19$.

To derive the absolute visual magnitude of the system, we calculated the
sum $M_V^{\rm pri} + M_V^{\rm sec} = -4.58$ with $M_V^{\rm pri} = -3.90$
for O9.5\,V and $M_V^{\rm sec} = -3.75$ for B0\,V; the latter value was
extrapolated from \citet{martins}. Eventually this leads to a distance
modulus of 10.36 or 10.77 for $R = 4.5$ or 3.1, respectively. The
corresponding distances of 1180\,pc and 1426\,pc display the influence of
the extinction law but nicely embrace the latest determinations for NGC\,6530.
Our estimate is also supported by the latest GAIA measurements
\citep{gaia2016} $d=1100\pm530$\,pc. Therefore, HD~164816 might well be a cluster member.

The rate of the apsidal advance is given as 0.67(7)~rad\,yr$^{-1}$ by
\citet{trepl}. A binary that has a similar spectral type and orbital
 period is Y~Cyg. Its apsidal motion is much slower, 0.1314~rad\,yr$^{-1}$
 \citep[see, e.g.][and references therein]{ycyg}; there are no obvious
 reasons for such a huge difference between those two binaries. Indeed,
although our spectra now cover a longer time interval than those used by
\citet{trepl}, we were unable to find any evidence for apsidal motion.

\begin{table*}
\caption{RVs based on the Gaussian fits of \ion{He}{i} line profiles of
HD~318015 (in \ks).}
\label{RV318}
\begin{tabular}{crrrrrrrrrrrc}
\hline\hline\noalign{\smallskip}
 RJD     &  Phase  &\multicolumn{3}{c}{6678 primary}&
\multicolumn{3}{c}{6678 secondary}&\multicolumn{2}{c}{ 5876 primary}&
\multicolumn{2}{c}{5876 secondary}&Spg.\\
         &         & RV      &error&\oc     & RV    &error&\oc     &RV&error&RV&error \\
\hline\noalign{\smallskip}
55721.7233&0.5107&  $-120$& 3&$-4.1$& $ 102$& 5&$ -1.7$& $ -130$ & 3& 107 & 8 & H      \\
55722.6926&0.5520&  $-124$& 3&$-3.0$& $ 104$& 5&$ -5.6$& $ -131$ & 3& 108 & 8 & H      \\
56100.6215&0.6712&  $-116$& 4&  1.3 & $  97$& 8&$ -8.3$& $ -139$ & 4& 111 & 7 & F      \\
56137.5210&0.2450&  --    &--&   -- &  --   &  &   --  & $  -45$ &  &$-19$&   & H      \\
56137.7109&0.2531&  --    &--&   -- &  --   &  &   --  & $  -41$ &  &$-23$&   & H      \\
56138.5066&0.2879&  --    &--&   -- &  --   &  &   --  & $  -41$ &  &$-35$&   & H      \\
56138.6052&0.2912&  --    &--&   -- &  --   &  &   --  & $  -41$ &  &$-35$&   & H      \\
56179.5219&0.0364&  $ 134$& 4&  2.3 & $-179$& 3&   2.5 &    129  & 3&$-176$& 4 & H      \\
56193.6246&0.6379&  $-124$& 4&$-2.6$& $  99$& 8&$-11.1$& $ -131$ & 5& 114 & 9 & F      \\
56376.8520&0.4580&  $ -92$& 5& 11.9 & $  85$&10&$ -4.9$& $  -97$ & 5& 107 & 9 & F      \\
56428.9173&0.6734&  $-108$& 5&  8.9 & $  98$& 8&$ -6.8$& $ -107$ & 4& 126 & 7 & F      \\
56517.5751&0.4548&  $ -97$& 4&  7.4 & $  98$& 7&   7.5 & $ -106$ & 3& 126 & 7 & F      \\
56520.7121&0.5886&  $-120$& 4&  2.0 & $ 119$& 7&   7.1 & $ -125$ & 3& 141 & 3 & F      \\
57131.7581&0.6505&  $-115$& 3&  5.2 & $ 106$& 4&$ -2.6$& $ -116$ & 3& 121 & 4 & C      \\
57132.7388&0.6923&  $-106$& 3&  7.1 & $  95$& 5&$ -5.5$& $ -113$ & 4& 100 & 9 & C      \\
57133.7728&0.7364&  $ -95$& 6&  5.0 & $  88$&10&   2.7 & $  -98$ & 4&  97 & 8 & C      \\
57134.8617&0.7828&  $ -77$& 7&  1.2 & $  82$&13&  21.7 & $  -82$ & 6&  59 &11 & C      \\
57135.8232&0.8238&  $ -42$& 7&  8.7 & $  41$&13&  12.4 & $  -51$ & 6&  31 &12 & C      \\
57136.6796&0.8604&  --    &--& --   &    -- &--&   --  &     --  &--&$-69$&-- & C      \\
57139.7835&0.9927&  $ 121$& 3&  2.4 & $-175$& 6&$ -8.6$&    114  & 3&$-161$& 4& C      \\
57143.7792&0.1632&  $  52$& 6&$-1.7$& $ -94$&11&$ -2.4$&     42  & 6&$-107$&10& C      \\
57143.8172&0.1648&  $  52$& 6&$-0.3$& $ -93$&11&$ -2.9$&     40  & 6&$-103$&10& C      \\
57145.8735&0.2525&  --    &--& --   &    -- &--&   --  &$  -36$  &  &$  -4$&--& C      \\
57146.7851&0.2914&  --    &--&  --  &    -- &--&   --  &$  -46$  &  &$  -6$&--& C      \\
57148.7816&0.3765&  $ -96$& 5&$-16.7$&   -- &--&   --  &$  -99$  & 9&$ -20$&11& C      \\
57150.7244&0.4594&  $-115$& 3&$-9.4$& $  93$& 5&   1.2 & $ -122$ & 4&   95 & 5& C      \\
57151.7681&0.5039&  $-124$& 3&$-9.2$& $  97$& 4&$ -5.5$& $ -137$ & 3&  106 & 4& C      \\
57152.8382&0.5495&  $-127$& 3&$-6.2$& $ 106$& 3&$ -3.4$& $ -141$ & 4&  129 & 5& C      \\
57154.7004&0.6290&  $-125$& 3&$-2.9$& $ 103$& 3&$ -7.9$& $ -131$ & 4&  104 &12& C      \\
57155.7111&0.6721&  $-125$& 3&$-7.9$& $ 103$& 4&$ -2.1$& $ -128$ & 4&  105 &12& C      \\
57156.7256&0.7153&  $-102$& 3&  5.0 & $ 100$& 4&   6.5 & $ -108$ & 4&  105 & 6& C      \\
57157.7168&0.7576&  $ -86$& 5&  5.1 & $  80$& 7&   4.8 & $  -85$ & 5&   92 & 9& C      \\
57158.7443&0.8014&  $ -66$& 7&  0.8 & $  64$&15&  16.9 & $  -71$ & 6&   35 &12& C      \\
57160.8764&0.8924&  --    &--&   -- &  --   &--&   --  &    --   &--&   13 &--& T      \\
57161.7600&0.9301&  $  42$& 7&$-16.3$&$-109$&13&$-12.1$&     34  & 6&$-125$&11& C      \\
57163.6970&0.0127&  $ 133$& 4&  4.7 & $-172$& 5&   5.6 &    132  & 3&$-164$& 4& C      \\
57164.6187&0.0520&  $ 133$& 5&  3.7 & $-172$& 7&   6.8 &    132  & 3&$-164$& 5& C      \\
57165.6736&0.0970&  $ 107$& 3&  0.4 & $-162$& 4&$-19.4$&    101  & 3&$-187$& 6& C      \\
57166.6550&0.1388&  $  71$& 4&$-3.0$& $-100$& 7&  15.0 &     69  & 4&$ -87$& 8& C      \\
\hline\noalign{\smallskip}
\end{tabular}
\end{table*}

\section{HD~318015 (V1082 Sco)}

This star belongs to the rare class of eclipsing binaries containing a~supergiant.
 It was classified as B0.7\,Ia by \citet{Massey}, B0.5\,Ib by \citet{Moffat} and
also independently by \citet{Drilling}. Its visual brightness is $V = 10$\m09 and
its variability was discovered by the Hipparcos satellite \citep{esa97}.

\smallskip\noindent
The orbital period and a linear ephemeris

\begin{equation}
T_{\rm min.I} = {\rm RJD}\, 52096.30 + 23\fd446
,\end{equation}

\noindent
was derived by \citet{Otero} from the Hipparcos \hp\ and ASAS~3 $V$ photometry.

There are seven FEROS and seven HARPS spectra in the ESO archive (exposed by
 Dr. Helminiak, with one exception: a FEROS spectrum at RJD~57160 was obtained
 in the service mode of the ``Tycho Brahe'' program). These reduced spectra
are available as pipeline products. As their phase distribution over the RV
curve contained gaps, we also obtained a series of R$\approx28000$ spectra
with the CHIRON spectrograph on the CTIO 1.5~m.
All RVs based on the Gaussian fits are listed in Table~\ref{RV318}. We also
measured RVs using \spefoe, but the results were nearly identical to those
derived via
Gaussian fitting, so we do not reproduce them here. The profile of the line
 \ion{He}{i}~6678~\AA\ at phase close to elongation is shown in
 Fig.~\ref{profile}. As V1082~Sco is heavily reddened
\citep[e.g.][$E_{B-V}=1.57$]{Moffat}, there are numerous diffuse interstellar
 bands in the spectra. Those around the line \ion{He}{i}~6678~\AA\ were also
observed in V1182~Aql \citep{1182}: 6660.5, 6672.4 and 6689.4~\AA, for example.

\begin{figure}
\resizebox{\hsize}{!}{\includegraphics{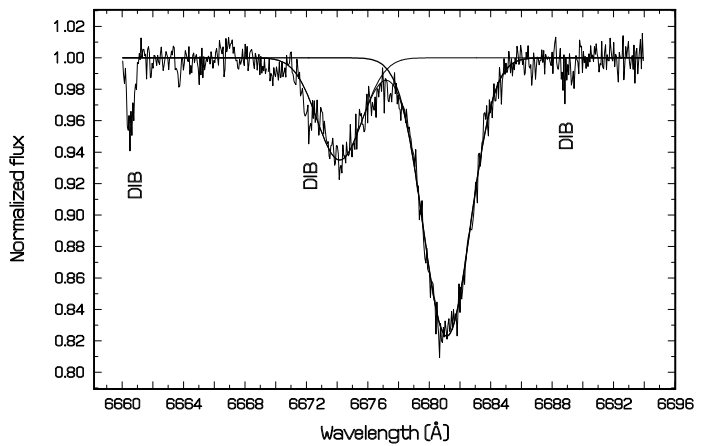}}
\caption{An example of \ion{He}{i}~6678 line profile at phase
near elongation of HD~318015.}
\label{profile}
\end{figure}

Although the measurements for the different \ion{He}{i} lines are not
identical, the RV curve is well defined and the solution leads to reliable
elements. They are listed in Table~\ref{orb318015} for the solution based on
RVs from the Gaussian fits of the \ion{He}{i}~6678~\AA\ line, with weights 1
for the primary, and 0.4 for the secondary RVs. The RVs close to conjunctions
 were not included in the solution.

\begin{table}
\begin{center}
\caption{Preliminary orbital elements for HD~318015 based solely on
the RVs from the Gaussian fits of the \ion{He}{i}~6678~\AA\ line;
same notation as in Table~3.}
\label{orb318015}
\begin{tabular}{lrccc}
\hline\hline\noalign{\smallskip}
Element                     & Value            \\
\hline\noalign{\smallskip}
$P$ [days]                  &  23.44598 fixed  \\
$T_{\rm periastr.}$ [RJD]   & 52075.91\,(15)   \\
$e$                         &     0.287\,(11)  \\
$\omega$ [$^\circ$]         &   343.7\,(2.6)   \\
$K_1$ [\ks]                 &   128.61\,(37)   \\
$K_2$ [\ks]                 &   147.47\,(40)   \\
$\gamma_1$ [\ks]           &  $-$28.8\,(1.3)  \\
$\gamma_2$ [\ks]           &     4.0\,(1.8)   \\
$m_1 \sin^3 i$ [\ms]        &    24.0\,(2)         \\
$m_2 \sin^3 i$ [\ms]        &    20.9\,(2)       \\
$a\,\sin i$ [\rs]           &   122.6\,(9)          \\
\vra [\ks]                    & 119\,(5)         \\
\vrb [\ks]                    & 104\,(8)         \\
rms$_1$  [\ks]                &     7.0          \\
rms$_2$  [\ks]                &     7.0          \\
\hline\noalign{\smallskip}
\end{tabular}
\end{center}
\end{table}

A solution of the Hipparcos and ASAS-3 light curves has already been obtained by
\citet{Zasche}, but the results had relatively large errors. As the orbital
eccentricity and the longitude of periastron are now also
better constrained with RVs, our combined light-curve (lc) and RV solution resulted in more
accurate elements (Table~\ref{lc318}); the fits of the ASAS and Hipparcos
light curves and of our RV curves are shown in Fig.~\ref{RVC_318}
(the RV curve based on the \ion{He}{i}~6678~\AA\ line was used since this line is free
of emission and represents the true orbital motion better than
the \ion{He}{i}~5876~\AA\ line). Our improved linear ephemeris is

\begin{equation}
T_{\rm min.I} = {\rm RJD}\, 53081.05 + 23\fd445975\,.
\end{equation}

The more massive star is larger and cooler, and its spectral lines are
considerably more pronounced. When parameters of the components are compared
with evolutionary models \citep[e.g. by][]{claret2004}, it appears that the age
of both components is approximately 6\,Myr.

We note that when the RV curves are solved independently from the light-curves
(see Tab.~\ref{orb318015}), slightly different values of the eccentricity and
the argument of periastron are obtained: $e=0.287\pm0.011$ and
$\omega=343.\!\!^\circ7\pm2.\!\!^\circ6$.
%though the difference is not large.
This might be the effect of circumbinary matter (CBM), similar to that which causes
non-zero eccentricity of RV curves of other early-type binaries
\citep{MD}.
Nevertheless our uncertainty estimates for the combined (light-curve and RV-curve)
solution are likely underestimated and the difference insignificant.

In the solution of light curves, one must adopt the effective temperature for one
of the components. In this particular case, the primary \teff\, is difficult to
estimate; for example, the \ion{He}{i} lines are almost insensitive to temperature
in the expected interval, and their EW, as well as that of other lines, are rather
large when compared to synthetic spectra. Fortunately, the effective temperature
of the less massive component is relatively well constrained. The \ion{He}{ii}~5411~\AA\
line is marginally present and the lines of the \ion{C}{ii}~6578 and 6583~\AA\ doublet
are absent. For \lgg $\sim$ 3.25, this restricts $T_{2,\rm eff}$ to
$29000 \pm 1000$\,K. The primary temperature, which follows from the \phoebe
solution is then compatible with the expectation for a B0.5/0.7 supergiant \citep{fraser,SK, mcerlean}.

\begin{table}
\caption{Combined \phoebe solution of the light and RV curves of HD~318015.
Error estimates are based on repeated runs of \phoebe with a~different
secondary effective temperature kept fixed throughout the fitting. All runs
together uniformly sample the interval $T_2 \in \left[28000; 30000\right]$\,K.}.
\label{lc318}
\begin{tabular}{lrrrrl}
\hline\hline\noalign{\smallskip}
Element                     & Value            \\
\hline\noalign{\smallskip}
Period $P$   [d]           &  23.445975(78)  \\
$T_{\rm periastr.}$ [RJD]  & 53083.80(16)    \\
$T_{\rm min.I}$ [RJD]      & 53081.05        \\
$T_{\rm min.II}$ [RJD]     & 53065.65        \\
Mass ratio $q$             & 0.861(5)       \\
Semimajor axis $a$  [\rs]  &  123.7(1.2)     \\
Eccentricity $e$           &   0.2675(12)    \\
Argument of periastron $\omega$ [$^\circ$] & 338.3(1.0) \\
Fract. radius $r_1$ [pole] &     0.2613(21)  \\
Fract. radius $r_2$ [pole] &     0.1371(11)  \\
Inclination [deg]          &    93.9(0.7)    \\
Radius $R_1$ [\rs]         &    33.39(55)       \\
Radius $R_2$ [\rs]         &    17.05(40)       \\
$T_1$ [K]                  & 24940(590)      \\
$T_2$ [K]                  & 29000(1000) \\
$m_1$ [\ms]                &    24.80(38)      \\
$m_2$ [\ms]                &    21.36(35)       \\
$L_1$ $[V]$                & 0.7490(33)      \\
$L_1$ $[H_{\rm p}]$        & 0.7447\,(32)      \\
$L_2/L_1$                  &     0.3427(62)      \\
$M_1^{\rm bol}$ [mag.]     &   $ -9.23$(1)        \\
$M_2^{\rm bol}$ [mag.]     &   $ -8.43$(2)        \\
log $g_1$   [cgs]          &     2.785(16)      \\
log $g_2$   [cgs]          &     3.304(14)      \\
Synchronicity ratio $F_1$  & 1.00 (assumed) \\
Synchronicity ratio $F_2$  & 1.00 (assumed) \\
\hline\noalign{\smallskip}
\end{tabular}
\end{table}

A disturbing fact is that the systemic velocity $\gamma$ of the primary is
more negative than that of the secondary by 35~\kms . Such a difference is often
present among early-type stars. A~very similar result has also been obtained
for another supergiant binary V1765~Cyg \citep{Hill,Ma91}, although in
this latter case, the detection of the secondary lines was uncertain; however,
in V1082~Sco the situation is extreme. A strong stellar wind of a binary
component is commonly considered as the reason for a blue shift of
its $V_\gamma$ velocity. It is known that the mass-loss rate by a radiatively
driven wind increases in proportion to the size of the star, and is especially
enhanced in the case of an evolved supergiant star. This is why the lines
of the primary are systematically blue-shifted with respect to those of
the secondary.

As the H$\alpha$ profile shows emission typical for an early supergiant
(see Fig.~\ref{HAL}), it is clear that some CBM must be present.
However, it is not easy to correlate the emission with either the primary
or the secondary star. We may note, however, that in some spectra the RV of
the emission peak follows the primary's RV, being more positive
by $\approx 150$~\ks.

In HD 318015, yet another uncommon characteristic is present: the depth
of the \ion{He}{i} lines of both components change severely,  ratio of depths
reaches 1:2 and the changes are only partly phase dependent.

 The estimated rotational velocities imply rotational periods of 14.1
and 8.0 days for the primary and secondary, respectively.
The expected \vsin for the synchronisation at periastron are 128~\kms
for the primary and 64~\kms for the secondary. In other words, the
primary seems to be synchronised at periastron while the secondary is rotating
almost twice as rapidly than what would correspond to synchronisation at
periastron.}

The parameters of the binary components can be compared with the evolutionary
models. We used the models by \citet{claret2004} and Padova
models\footnote{Available at http://stev.oapd.inaf.it/cgi-bin/cmd}
 \citep{bressan2012,rosenfield2016} for masses of 20 to 30~\ms.
 The Claret models are shown in Fig.~\ref{evol}; the fit is poor,
and is not improved when the Padova models are used. Although the mass of
the donor is commonly accepted as higher, in this case both components are
over-luminous for their masses. In order to make the agreement better, the
components would have to be more massive, larger, or cooler. As the masses
and radii appear as well determined, the temperature might be suspected,
although the necessary difference of $\approx 4000$ K is not compatible
with the characteristics of the spectra.

We note that comparisons of OB supergiant spectra with line-blanketed
model atmospheres and determination of their physical properties was
carried out by \citet{crow} and \citet{fraser}, for example.
In Fig.~\ref{crow}, the properties of the supergiants studied by \citet{crow}
and those of the components of V1082~Sco
are compared with the evolutionary tracks by \citet{schaerer93, schaller93}.
The primary component fits the line defined by the other spectra perfectly;
note, however, that the stars shown there have masses from 14 to 46~\ms\, and
do not indicate any obvious dependence of the mass on the effective
temperature. The spectra studied by \citet{fraser} indicate a similar
behaviour; see their Fig. 3. Although they occupy a correct area,
the fit with masses does not exist.

   We underline that in the above exercises,
we deliberately used several different grids of stellar evolutionary models
to demonstrate that the discrepancy is inherent to the systems and
does not depend on the choice of a particular set of evolutionary models.
Using common procedures, the distance of the binary appears to be 2800 pc.
The binary is a member of the open cluster Trumpler 27. Although there are
doubts about the reality of the cluster, several similar early-type stars
in the vicinity of V1082~Sco are certain \citep{Moffat}.

 In passing we note that the only eclipsing binary in our sample,
V1082~Sco, can be compared to V1176~Cen, eclipses of which were discovered
by \citet{otero2005}, who used the $V$ observations from the ASAS3
survey; the period of V1176~Cen is 31\fd029. Solving this ASAS3 light curve
with \phoebee, we found relative radii of 0.154 and 0.0724, and
an inclination of $87^\circ9$. If $T_1=22000$~K is assumed, then
$T_2=17600$K.

\begin{figure}
\resizebox{\hsize}{!}{\includegraphics{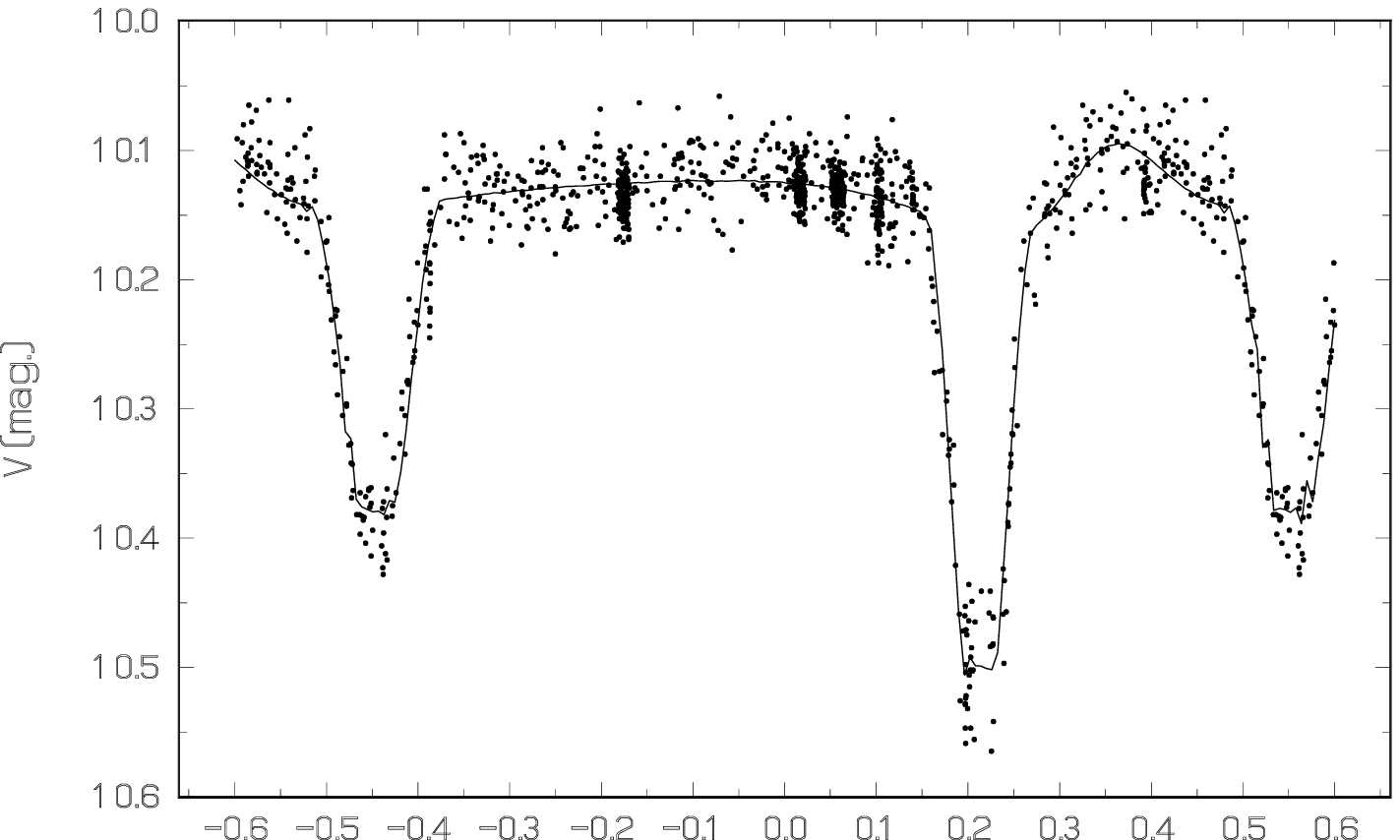}}
\resizebox{\hsize}{!}{\includegraphics{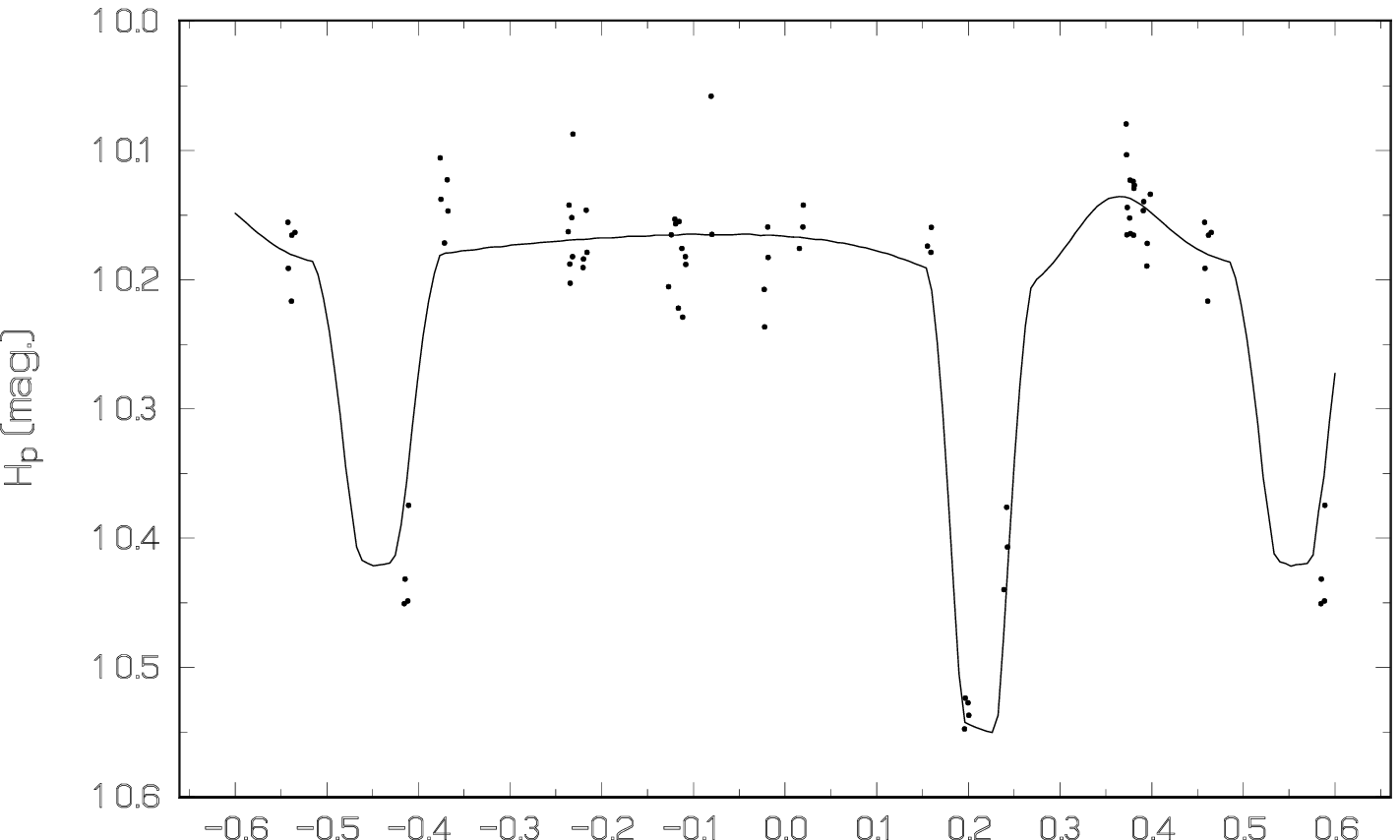}}
\resizebox{\hsize}{!}{\includegraphics{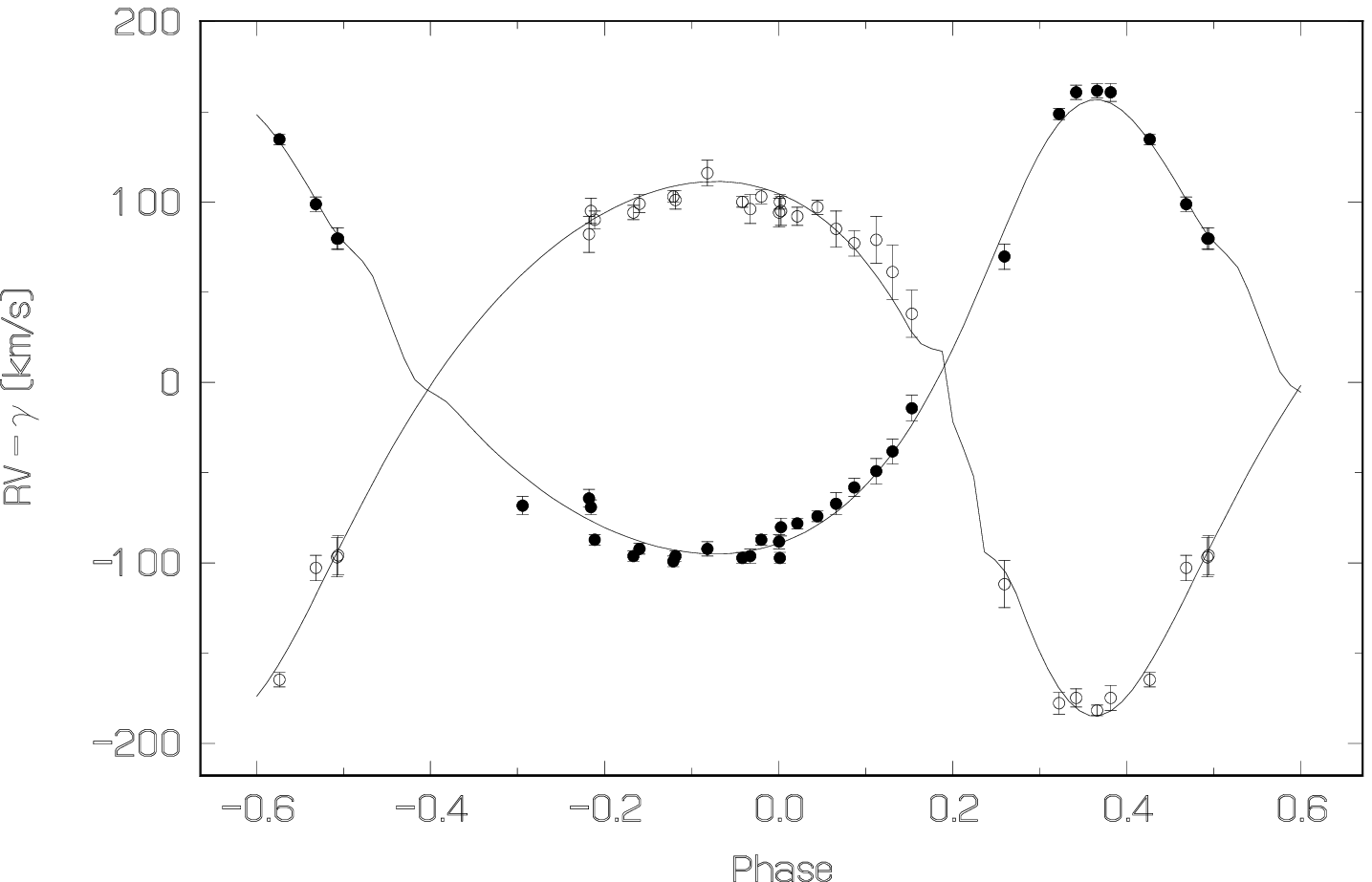}}
\caption{The ASAS3 $V$ and Hipparcos \hp\ light curves and the orbital
RV curve of HD~318015 based on RVs from the Gaussian fits of
the \ion{He}{i}~6678~\AA\ line. Phase zero corresponds to RJD 53076.08,
chosen as an arbitrary reference epoch in \phoebee. The phases of the
primary and secondary minima according to the combined \phoebe solution
are then 0.2120 and $-0.4449$, respectively.}
\label{RVC_318}
\end{figure}

\begin{figure}
\resizebox{\hsize}{!}{\includegraphics{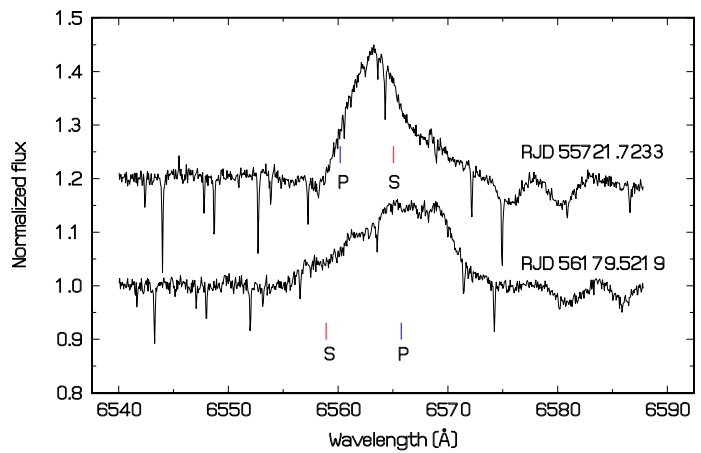}}
\caption{Two examples of H$\alpha$ line profiles of HD~318015.
The \ion{C}{ii} doublet at 6578 and 6582~\AA\ is also seen.
The vertical dashes show the position of the primary (blue) and the
secondary (red) in both \ha profiles.}
\label{HAL}
\end{figure}

\begin{figure}
\resizebox{\hsize}{!}{\includegraphics{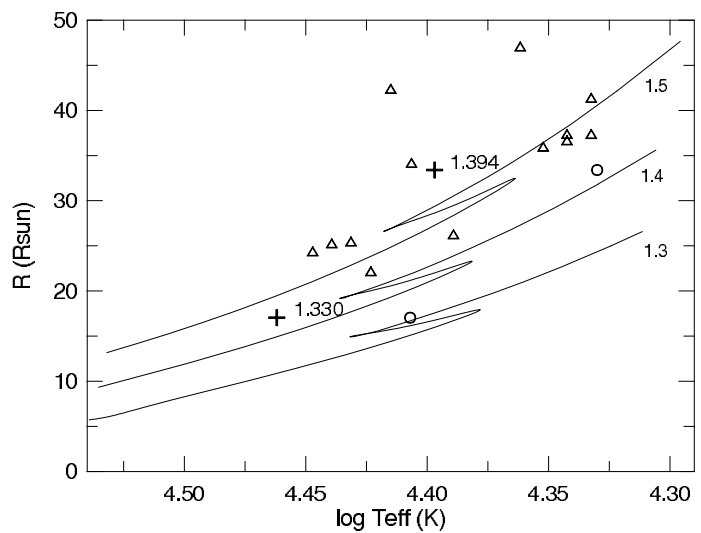}}
\caption{HD~318015: the lines show the relation between $ \log T_{\rm eff}$
and $R$ for the evolutionary models published by \citet{claret2004}. The
lines are labelled by $\log M$. Two plus symbols are the positions of both
binary components, also labelled by $\log M$. Two circles show where the
stars with the given masses and radii should
be located.}
\label{evol}
\end{figure}

\begin{figure}
\resizebox{\hsize}{!}{\includegraphics{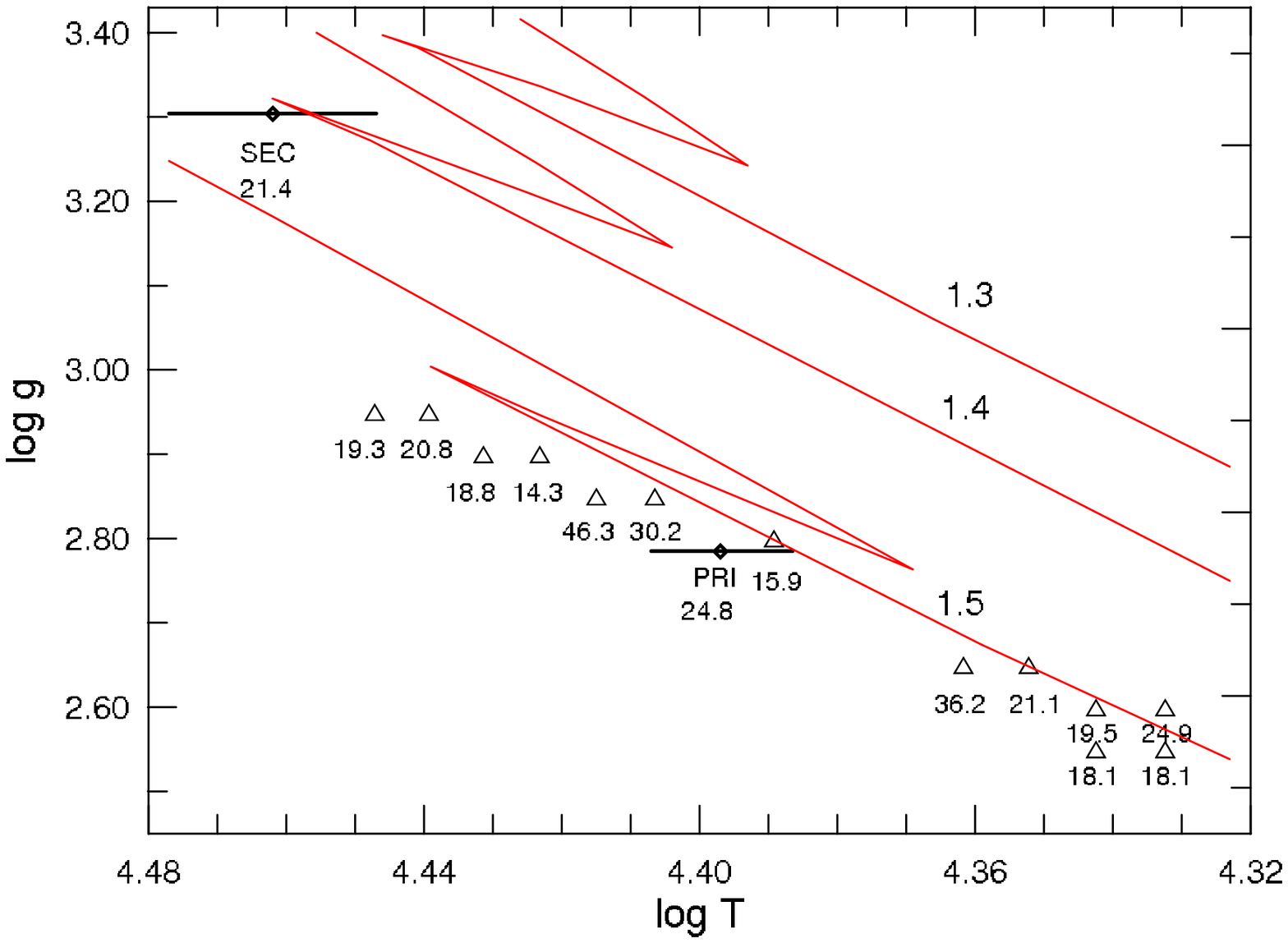}}
\caption{A comparison of the properties of the components of V1082~Sco
with those for OB supergiants studied by \citet{crow} in the $\log~g$
vs. log~\tef diagram. Evolutionary tracks from \citet{schaerer93} are also
shown. Note that there is no clear correspondence between the \tef
and stellar masses.}
\label{crow}
\end{figure}

\section{Conclusions}

Studying seven high-mass objects, we found that a detailed investigation
has increased the multiplicity status of each system,  be it from single to
binary or from binary to triple. Table~\ref{Multiplicity} shows that
the secondary components are O and B-type stars.
In at least two cases, the tertiary component is also a massive star.
Although the sample is small, the number of multiple systems is
surprisingly high. Five out of seven systems studied here contain
a~third component or are suspected to have one.

The two short-period $P_{\rm inner} < 7$\,d systems (HD~92206~C and
HD~123056) were both identified as (at least) triple systems in
agreement with statistical analysis by~\citet{tokovinin2006}, who
reported
that 96\% of binaries with a period of less than three days are at
least triple systems. The multiplicity fraction among the remaining
long-periodic ($P_{\rm inner} > 7$\,d) binaries of $\approx 50\,\%$
also agrees with their findings. It is difficult
to discuss the formation process of the individual multiples, because with
the exception of HD~123056, the properties of the outer orbit are
unknown. Nonetheless, the members of the inner binaries of both
short-period systems have close-to-equal masses, which is
a~typical outcome of accretion from a~circumbinary disk
\citep{bate2002}. The inner and outer orbits of HD~123056
are likely close enough for the Kozai-cycle with tides
\citep{eggleton2001,fabrycky2007} to play some role in the
shrinkage of its inner orbit. However, without knowing the orbital elements
of both orbits, it is impossible to (dis)prove such a conjecture.

{Our results for \vsin values seem to confirm the conclusion of
\citet{ramirez} about the lack of very high rotational velocities
among O-type binary systems, at least for cases where we have some idea
about the orbital inclinations.}

\begin{table}
\begin{center}
\caption{Multiplicity properties of the new OB systems}.
\label{Multiplicity}
\begin{tabular}{lcccc}
\hline\hline\noalign{\smallskip}
         \multicolumn{1}{c}{HD}
       & \multicolumn{1}{c}{primary}
       & \multicolumn{1}{c}{secondary}
       & \multicolumn{1}{c}{$m_2 / m_1$}
       & \multicolumn{1}{c}{tertiary}\\
\hline\noalign{\smallskip}
~\,92206\,C& O8\,V      & O9.7\,V   &    0.75       &   B2:    \\
~\,93146\,A& O6.5\,V    & B0.5      &    0.48       & O9.7\,IV$^1$ \\
112244     & O8.5\,Iab  &     late B?     & $0.06 - 0.12$ & unknown  \\
123056     & O8.5       & B0:       &    0.70       &  B0:     \\
164438     & O9.2\,IV   & later B3  & $0.10 - 0.22$ & probable \\
164816     & O9.5\,V    & B0\,V     &    0.90       &   yes    \\
318015     & B0.7\,Ia   & O9.5\,III &    0.81       & unknown  \\
\hline\noalign{\smallskip}
\end{tabular}
\end{center}
\footnotesize{Notes. Spectral types marked by a colon (:) are uncertain. $^1$The
physical association of this nearby star needs to be verified.}
\end{table}

 Our general conclusion is that the ultimate decision among various theories
of the massive-star origin might only come when the binary properties of
a~representative sample of systems, based on individual orbital solutions,
and the light-curve solutions in case of eclipsing (sub)systems,
are known. So far, the knowledge of duplicity and/or multiplicity of
OB stars has mainly been deduced using statistics.

Furthermore, we note that the rapidly maturing optical interferometry can also
significantly contribute to achieving the above-mentioned goal.
For instance, the primary of HD~93146~A is the earliest star in our sample and
has a long period of 1131.4\,d. Both this star and HD~123056, another star
that probably has a period of 1314\,d, might be suitable
targets for interferometry. HD~123056 was already observed by
\citet{aldo2015}, but not separated; higher resolution than
10\,mas is probably needed. Also in the case of HD~93146~A, high
resolution is necessary; at a distance of $\approx 3.0$\,kpc its
semi-major axis might  be of the order of 2.5\,mas.

\begin{acknowledgements}
This publication is supported as a project of the Nordrhein-Westf\"{a}lische
Akademie der Wissenschaften und der K\"unste in the framework of the academy
programme by the Federal Republic of Germany and the state Nordrhein-Westfalen.
In the early stages of this project, the research of PH, PM, and JN was supported
by the grant P209/10/0715 of the Czech Science Foundation. JN and PH were also
supported from the grant GA15-02112S of the Czech Science Foundation and
from the grant No. 250015 of the Grant Agency of the Charles University in
Prague.
We are grateful for the help of A. Barr Dom\'{i}nguez, K. Fuhrmann,
L. Kaderhandt and M. Ramolla during the observations and the reduction.
We also thank High Point University for providing the funds necessary to
purchase observing time on the CTIO 1.5~m telescope through the SMARTS
Consortium, and for funding part of this work during the 2015 HPU
Summer Research Program in the Sciences (SuRPS).
The observations obtained with the MPG 2.2~m telescope were supported by
the Czech Ministry of Education, Youth and Sports project LG14013 ("Tycho Brahe:
supporting Ground-based Astronomical Observations") during run P2
in May 2015. We would like to thank the observers Drs. P. Kabath and
S. Vennes for obtaining the data.
We thank Universidad Cat\'{o}lica del Norte in Antofagasta, Chile, for
continuous support.
 Our thanks are also due to Dr.~M.~Hackstein, who put at our disposal
the individual photometric observations of HD~92206C.
The use of the NASA/ADS bibliographical service and SIMBAD electronic
database are gratefully acknowledged.
\end{acknowledgements}

%\vfill\eject
%%HD References were not checked by me
\bibliographystyle{aa}
\bibliography{hd}

\end{document}